\documentclass[twocolumn,aps,showpacs,prb,epsf,graphics,psfig]{revtex4}
\usepackage{graphicx}
\usepackage{graphicx}
\usepackage{bbold}
\usepackage{color}
\usepackage[normalem]{ulem}
\usepackage{amssymb}

\begin{document}
\title{
Renormalization of radiobiological response functions by energy loss fluctuations and complexities in chromosome aberration induction: deactivation theory for proton therapy from cells to tumor control
}

\author{Ramin Abolfath$^{1,2,3,\dagger}$, Yusuf Helo$^{4}$, Lawrence Bronk$^1$, Alejandro Carabe$^2$, David Grosshans$^1$, Radhe Mohan$^{1,*}$}
\affiliation{
$^1$Department of Radiation Physics and Oncology, University of Texas MD Anderson Cancer Center, Houston, TX, 75031, USA \\
$^2$Department of Radiation Oncology, University of Pennsylvania, Philadelphia, PA 19104, USA \\
$^3$Department of Therapeutic Radiology, Yale University School of Medicine, New Haven, CT 06520, USA \\
$^4$Imanova Centre for Imaging Sciences, Invicro LLC, London, United Kingdom
}

%

\date{\today}

\begin{abstract}
We employ a multi-scale mechanistic approach built upon our recent phenomenological / computational methodologies (Ref. [\onlinecite{Abolfath2017:SR}]) to investigate radiation induced cell toxicities and deactivation mechanisms as a function of linear energy transfer in hadron therapy.
Our theoretical model consists of a system of Markov chains in microscopic and macroscopic spatio-temporal landscapes, i.e.,
stochastic birth-death processes of cells in millimeter-scale colonies that incorporates a coarse-grained driving force to account for microscopic radiation induced damage.
The coupling, hence the driving force in this process, stems from a nano-meter scale radiation induced DNA damage that incorporates the enzymatic end-joining repair and mis-repair mechanisms.
We use this model for global fitting of the high-throughput and high accuracy clonogenic cell-survival data acquired under exposure of the therapeutic scanned proton beams, the experimental design that considers $\gamma$-H2AX as the biological endpoint and exhibits maximum observed achievable dose and LET, beyond which the majority of the cells undergo collective biological deactivation processes.
An estimate to optimal dose and LET calculated from tumor control probability by extension to $~ 10^6$ cells per $mm$-size voxels is presented.
We attribute the increase in degree of complexity in chromosome aberration to variabilities in the observed biological responses as the beam linear energy transfer (LET) increases, and verify consistency of the predicted cell death probability with the in-vitro cell survival assay of approximately 100 non-small cell lung cancer (NSCLC) cells.
The present model provides an interesting interpretation to variabilities in $\alpha$ and $\beta$ indices via perturbative expansion of the cell survival fraction (SF) in terms of specific and lineal energies, $z$ and $y$, corresponding to continuous transitions in pair-wise to ternary, quaternary and more complex recombination of broken chromosomes from the entrance to the end of the range of proton beam.
\end{abstract}
\maketitle
\section{Introduction}
The Bragg peak and capability in delivering a sharp dose deposition pattern in deep-seated cancerous / malignant tumors is the main advantage of using protons and Heavy ions in radiation therapy~[\onlinecite{HallBook,Olsen2007:RO,Schardt2010:RMP,Durante2011:RPP}]. However, a complete and robust model that describes the cellular response and the biological effectiveness of the charged particle along the track of radiation beam is still lacking.

In particular, uncertainties in the biological data at locations distal to the Bragg-peak (just beyond the Bragg peak, i.e., in the tail region) where the charged particle stopping power diverges are the main challenge for the biological dose calculation in treatment planning systems. Thus, open questions on the biological effect of particles, which is often expressed in terms of relative biological effectiveness to photons~[\onlinecite{MacDonald2006:CI}] (RBE), are yet to be answered.

It is known that generated ions and electrons (products of ionizations and not e.g. recoil nuclei) along the path of a particle track crossing a DNA material can potentially induce double strand breaks (DSBs)~[\onlinecite{HallBook,Schardt2010:RMP,Durante2011:RPP}].
The DNA damage and the subsequent biochemical pathways~[\onlinecite{43}] induced by ionization
and non-ionizing ($meV$) molecular excitations, including thermal conduction and shock waves were recently investigated~[\onlinecite{Toulemonde2009:PRE,Surdutovich2010:PRE,Solovyov2017:Book,Surdutovich2014:EPJD,Verkhovtsev2015:SR}].

The DSBs are fundamental elements in causing lethal damages to cells and their mitotic / meiotic offsprings.
Following DSB formation, enzymatic repair processes trigger chromosomes end-joining, either through homologous or non-homologous pathways, depending upon the cell type and cell cycle.
This process determines the particle RBE and is error-prone meaning that there are possibilities in mis-joining chromosomes that may yield apoptotic or necrotic cell death~[\onlinecite{HallBook}] and may increase the risks of second malignancies~[\onlinecite{Kohandel2015:IJRB}].
In general, RBE depends on the fractionation scheme, biological endpoint, radiation quality, the tissue and cell-type.

Recently a graded solid water compensator was designed to allow irradiation of cells by mono-energetic scanning beam of protons at specific depths~[\onlinecite{Guan2015:SR}]. Subsequent high-throughput automated clonogenic survival assays were performed on non-small cell lung cancer (NSCLC) cells to spatially map the biologic effectiveness of scanned proton beams with high accuracy.

This method reduces uncertainties in biological data. However, the current RBE models~[\onlinecite{Butts1967:RR,Katz1971:RR,Kraft2000:NP,Paganetti2001:IJRB,Elsasser2010:IJROBP,Hawkins1998:MP,Hawkins2003:RR,Kase2006,22a,23a,24a,25a,27a,Wilkens2004:PMB,Neary1965:IJRB,Steinstrater2015:PMB,McNamara2015:PMB,Curtis1986:RR,Sachs1997:IJRB,Steinstrater2012:IJROBP,Carlson2008:RR,Frese2012,Stewart2015}] including
the microdosimetric kinetic model (MKM)~[\onlinecite{Hawkins1998:MP,Hawkins2003:RR,Kase2006}], local effect model (LEM)~[\onlinecite{22a,23a,24a,25a,27a}], Wilkens and Oelfke [\onlinecite{Wilkens2004:PMB}], Steinstr\"ater {\em et al.} [\onlinecite{Steinstrater2012:IJROBP}], and Monte Carlo damage simulation (MCDS-RMF)~[\onlinecite{Carlson2008:RR,Frese2012,Stewart2015}],
do not adequately explain these data. In particular, these models underestimate the increase in radiobiological response functions, i.e., linear-quadratic (LQ) $\alpha$ and $\beta$ indices, in addition to RBE in locations beyond the Bragg-peak.

Fitting of several existing methods to the experimental data has recently investigated and presented in Ref.[\onlinecite{Mohan2017:ARO}].
Accordingly, these models are not capable of explaining the high non-linearity of RBE that was observed in pre-clinical studies at higher LET at points beyond the Bragg peak.
An increased complexity in double strand breaks, that is the focus of the present work, was suggested to attribute to the experimentally observed non-linearity at higher LETs.

Recently a microscopic computational study based on interfacing Monte Carlo track structure calculation and DNA molecular dynamics (MD)~[\onlinecite{Abolfath2011:JPC,43,Abolfath2016:MP}], predicted the increase in relative DSBs induction of protons up to a factor of 4 compare to a beam of radiation with LET of unity. 
This is consistent with the RBE measurement, reported by Guan {\em et al.}~[\onlinecite{Guan2015:SR}].
In addition, effects of low-energy protons and ions have been simulated in detail recently by Friedland {\rm et al.}~[\onlinecite{Friedland2017:SR}] and Meylan  {\rm et al.}~[\onlinecite{Meylan2017:SR}].
The predicted enhancements in yields of DSB and DNA fragments with increasing LET is qualitatively comparable to the quoted number and data in Ref.~[\onlinecite{Guan2015:SR}].

Recently a computational model for relative biological effectiveness of therapeutic proton beams based on a global fit of cell survival data has been introduced by our group in Ref.~[\onlinecite{Abolfath2017:SR}]
where a numerical approach in fitting a surface to survival fraction (SF) experimental data in a three-dimensional parameter space using two independent variables, dose and linear energy transfer (LET), has been implemented.
This approach allows achieving a smooth relation between cell survival and LET for the entire range, of LET up to the end of the beam range where LET grows non-linearly before it vanishes.
The improvement in the numerical stability of the extracted RBE data has revealed the superiority against the traditional fitting approaches [\onlinecite{HallBook}] where the fitting is performed for each individual survival curve with a specific average LET.
We carried out an iterative “global” fit to the measured data and calculated RBE self-consistently and showed that the process reduces the overall uncertainty.
The results in fitting SF's for two separate cell lines, H460, and H1437 show a non-linear increase in RBE in domains distal to the Bragg peak.
Simultaneous fits of SF data in dependence on the dose and LET which have been examined and presented by Kundrat {\em et al.}~[\onlinecite{Kundrat2005:PMB,Kundrat2006:PMB}].

In spite of reliability and success in numerical results, it would be challenging to quantitatively identify the biophysical processes throughout the fitting data.
Particularly in practical approaches, the LET dependence of cell-survival data may vary among different fitting procedures in a range of the beam where experimental uncertainties are significant, i.e.,
by the end of the proton range, distal to Bragg peak, where measuring the dose deposition and identifying the accurate value of LET are challenging.

To address these issues, we developed an analytical analogue and present a theoretical approach and mathematical details of the fitting model.
In particular we present a detail description in the relative biological effectiveness of proton beams as a function of their LET. The model aims at capturing the stochastic nature of energy deposition as well as complex patterns of chromosome aberrations.
The present model relies on microdosimetry and its formalism, a model formulated based on generalization of MKM.
Event-by-event track structure simulations as described by Nikjoo {\em et al.} [\onlinecite{Nikjoo2016:RPP}] and Friedland {\em et al.} [\onlinecite{Friedland2011:MR}] were performed to capture the energy deposition and subsequent processes in greater detail.

In addition to the initial damage clustering, the chromatin dynamics and the resulting mobility of induced intra- and inter-chromatin ends have been incorporated in our model through phenomenological repair and mis-repair rate equations. The latter accounts for the exchange of chromosome fragments and rejoining of different chromosome ends among each other. Numerous studies on aberration formation have estimated that only breaks within about 1 micrometer may misrejoin and form aberrations (e.g., see for example Refs. [\onlinecite{Sachs1997:RR,Ballarini2004:CGR}].
Track structure-based simulations have been extended to DSB repair and formation of chromosome aberrations. Both temporal and spatial effects have been represented via interplay of enzymatic processing of DNA termini and their mobility (e.g., see for example Refs. [\onlinecite{Friedland2010:RR,Friedland2013:MR}]).

The present model predicts interesting phenomenon, i.e., occurrence of continuous transitions in population of chromosome aberration complexes amongst binary, ternary, quaternary and higher order combinations as LET of scanned beam varies continuously.
The predictions and mathematical hypothesis of evolution of chromosome complexities as a function of LET has not been verified experimentally. However, we propose a systematic experiment to be performed to investigate the complexity of chromosome aberration as a function of particle LET.
Such measurement can be fine-tuned to perform biological spectroscopy of the beam energy loss and to be considered as a signature of particle RBE.

We devote last sections to present calculation of the lethal lesions and tumor control probability (TCP) as a function of dose and LET for non-small cell lung cancer (NSCLC), H460 and H1437. The framework of our computational platform is based on generalized MKM and a multi-scale three-dimensional global fitting~[\onlinecite{Abolfath2017:SR}] of the cell survival, recently measured by our experimental group in MD Anderson~[\onlinecite{Guan2015:SR}].
Our methodology is consistent with other methodologies for proton and ion beam therapy, based on a multiscale framework that have been developed and applied successfully to describe cellular response and RBE for ion irradiation (see for example Ref.~[\onlinecite{Surdutovich2010:PRE,Solovyov2017:Book}] and references therein).
We further verify matching of the cell death probability calculated specifically for the in-vitro cell survival assay with a maximum achievable dose and LET observed in the experiment, beyond which the majority of the cells exhibit biological deactivation. We finally summarize our work with discussion and conclusion.

\section{Materials and Methods}
\subsection{Basic definitions and model calculation}
\label{sect_1}
Before going through the details and mathematical construction of the present model, we first clarify the definition of ``event'' consistent with the collective phenomenon and formulation of DSBs developed in this work, and non-Poissonian distribution of ionizations based on Kellerer-Rossi's theory of dual radiation action~[\onlinecite{Kellerer_Rossi1972:CTRR}].
In this context, ``an event'' refers to a series of energy depositions in ionization processes induced from passage of ``a primary charged particle'', hence
an event refers to all spatiotemporal energy-depositions induced by passage of a single proton through a volume of interest, e.g., a cell nucleus.

Occurrence of an individual ionization takes place within an electrodynamic time scale ($10^{-18}$-$10^{-17}$ s), and in subatomic length scales. The time and length scales are fundamental / natural scales relevant to the atomistic excitations.
Traversing a primary charged particle through a medium creates a collection of spatially scattered ionizations and release of secondary charge particles, all can be mapped into a particle's entity, known as a single particle track structure.
Hence we refer a single-event scored in DNA material to a single track structure generated from passage of a primary charged particle.

\begin{figure}
\begin{center}
\includegraphics[width=0.8\linewidth]{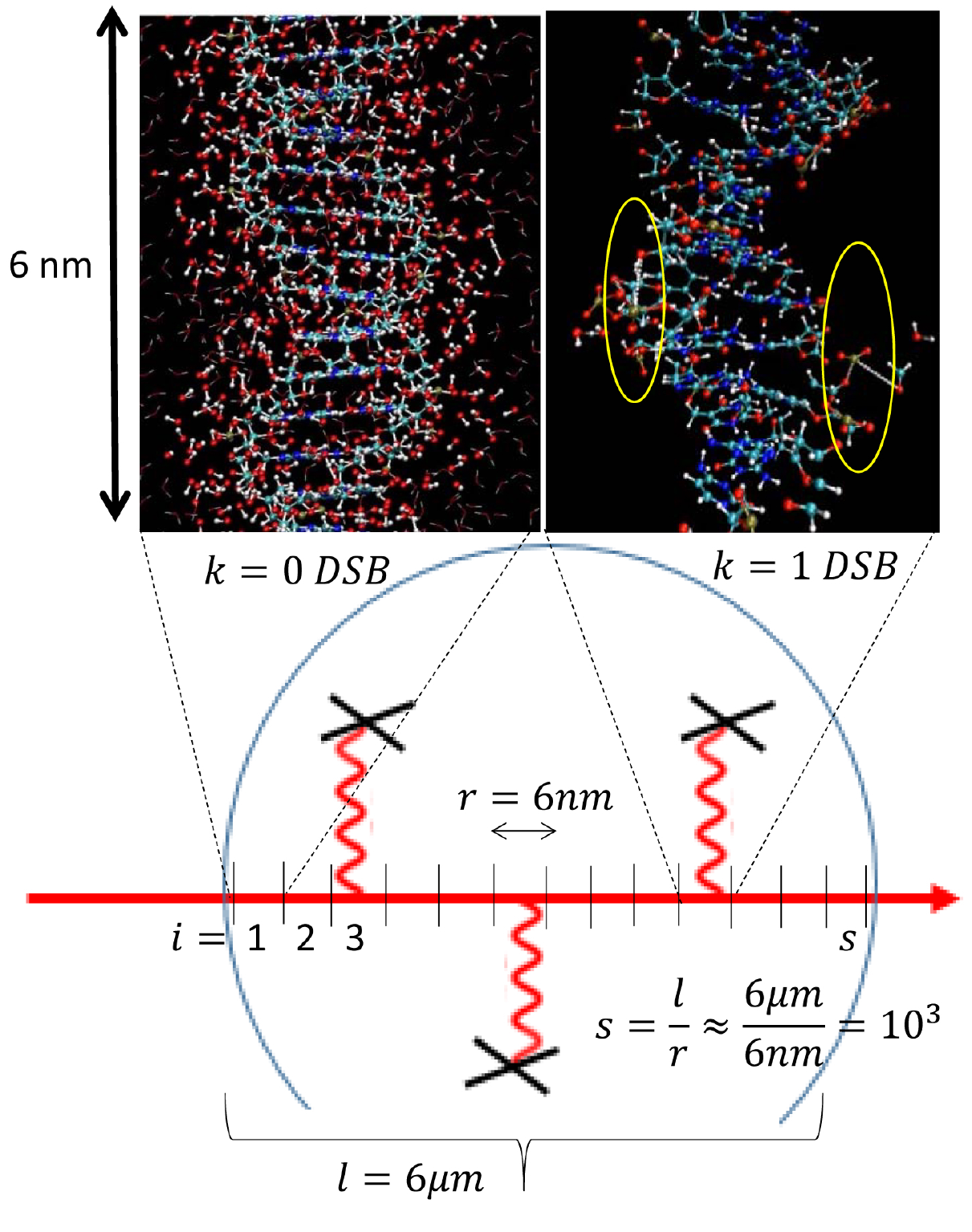}\\ 
\noindent
\caption{
Schematically shown the partitioning of a cell nucleus (circular structure) into segments of proton-tracks (red lines) used as unit of ionization per MD simulation of DNA damage.
The model calculation starts with scoring the ionizations and the time-evolution of chemical reactivity of species induced by ionized water molecules, in surrounding of DNA in sub nanometer and femto-second spatiotemporal scales.
In upper left corner, a magnified structure of DNA surrounded by water molecules is depicted with no scored DSB induction.
Shown in upper right corner, a typical snap-shot of distorted DNA obtained after running MD for $\approx$ 50 ps. Two SSBs are highlighted by yellow circles, located within 10 base-pairs separation are indication of a DSB formation. Damaged bases, distorted hydrogen bonds, base-stacking as well as complex species such as hydrogen-peroxides formed from two OH$^.$ free radicals are visible.
For clarity of the visualization, we removed water molecules from the image of the DNA molecule.
}
\label{Fig000x}
\end{center}\vspace{-0.5cm}
\end{figure}

\begin{figure}
\begin{center}
\includegraphics[width=0.8\linewidth]{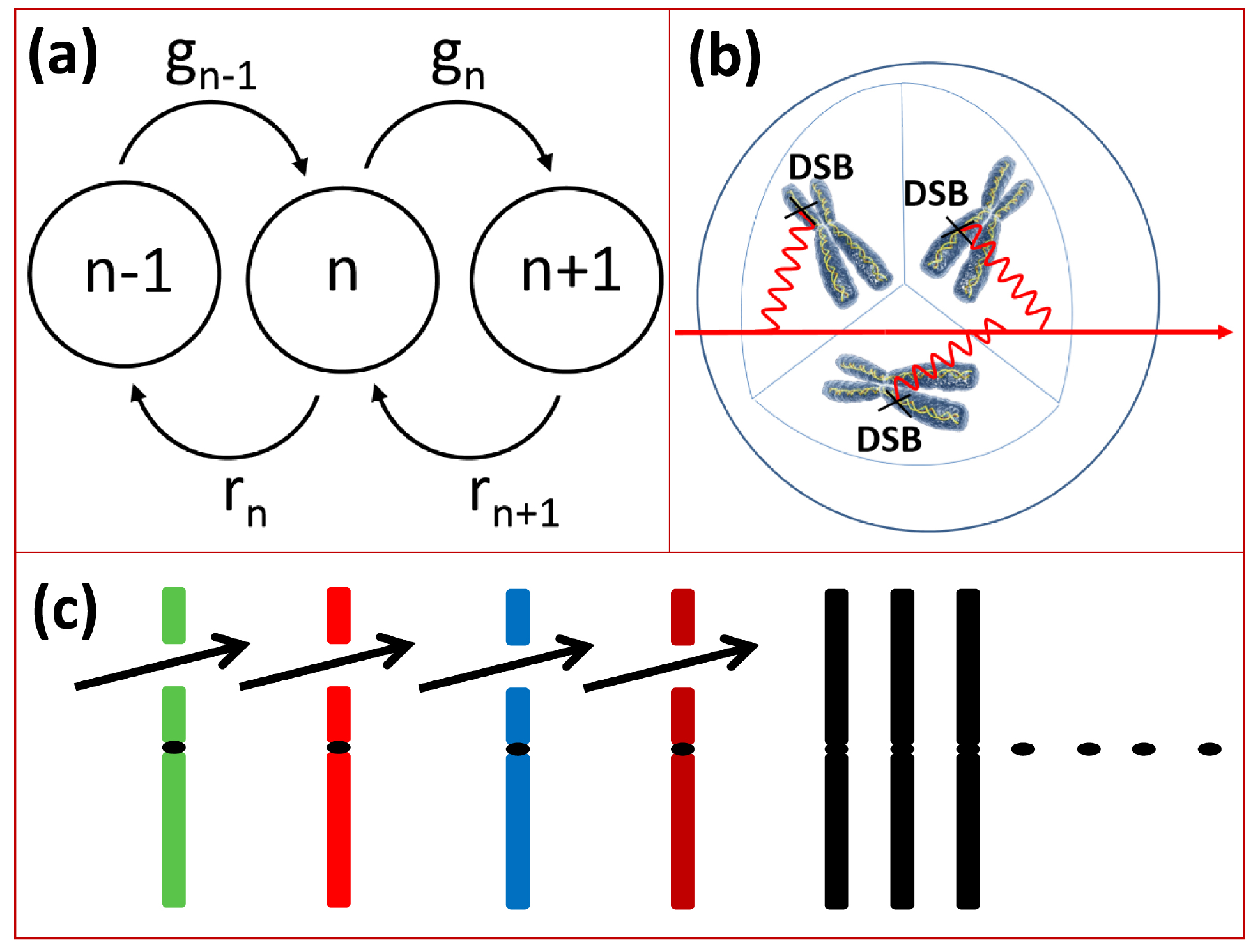}\\ 
\noindent
\caption{(a) Schematic diagram of Markov process in DSB rate equation, Eq.~(\ref{eq002}).
(b) Schematic representation of DSB induction in a cell nucleus.
The bold arrow represent a charged particle traversing cell nucleus. The wiggly lines were adopted from Feynman's diagrams in quantum electrodynamics (QED) to describe propagation of interaction of a particle as a field in scattering processes with an interaction-site (see for example Ref.~[\onlinecite{Mandle_Shaw_book}]).
(c) Chromosomes undergo DSB induction after track of particles traverse the cell. Colored lines represent chromosomes and the gap between each chromosome represent a DSB. The black dots in the middle of each chromosome represent centromere. Black chromosomes were not gone through DSB formation after traversing charged particles.
}
\label{Fig00}
\end{center}\vspace{-0.5cm}
\end{figure}

Out of large number of energy deposition processes along a single track, few nm-scale DSBs form in DNA materials in nm-size target volumes.
This number depends on the size of ionization clusters and the energy balance requirement needed for induction of DSBs.
Figures ~\ref{Fig000x} and ~\ref{Fig00} schematically illustrates this process.

Scoring the processes that initially yield DNA hydrogen abstraction, and by several orders of magnitude elapses to DSB formation, cell lethality and tissue toxicities, requires modeling in multi-scale landscapes.
Although the basis of multi-scaling in our approach is conceptually consistent with similar studies recently reported in the literature, (e.g., see Ref.[\onlinecite{Surdutovich2010:PRE,Solovyov2017:Book,Surdutovich2014:EPJD}]), but there are differences that require to be clarified.

Our computational steps begin with simulation of the initial physical and chemical DNA damage processes in atomic scale, as shown in Figure \ref{Fig000x}. In this step, we first score energy and coordinates of all types of molecular excitations, in addition to atomic ionizations using Geant4-DNA~[\onlinecite{Incerti2010:IJMSSC}].
Because of significant difference in time scales between the direct energy transfers to water molecules surrounding DNA and indirect energy transfers to DNA via chemical reactions and free radicals, in the second step of computation, the ionization coordinates, energy transfers to medium, stepping length of each interaction, the distance the particle travels before losing energy by interaction, and other useful information obtained from Geant4-DNA, will be cast into a look-up table to be used as initial conditions to a ReaxFF MD, a quantum mechanical model appropriate for simulation of chemical reactions. The ReaxFF MD simulation box / voxel consists of a double-strand DNA molecule, water molecules that form an aqueous environment in addition to free radicals that invade the DNA molecule and perform hydrogen abstraction.

The details of these steps, originally presented in Refs. [\onlinecite{Abolfath2011:JPC,43}], are schematically illustrated in Figure \ref{Fig000x} where
a red arrow shows a single track-structure of a proton traversing a cell nucleus (circular structure). We divide the cell nucleus into nm-size domains based on the path length of the proton track and the dimension of simulation box / voxel. Hence the size and number of domains are determined by diameter of cell nucleus, the length of track-structure and the size of simulation box used in ReaxFF MD. An example of such configuration is sketched in Figure \ref{Fig000x} where a cell nucleus with diameter 6$\mu m$ is depicted. Furthermore, in order to score at least one DSB, a DNA with at least 10 base pairs must be constructed in ReaxFF MD computational box. This requires fitting of a DNA structure with an approximate length greater than 6 nm.
Hence the track and cell nucleus along the path of proton must be partitioned into approximately 1000 segments, as depicted in Figure \ref{Fig000x}.

In each segment, we score number of DSBs using a selective statistical sampling algorithm introduced in Ref.[\onlinecite{43}]. Summing DSBs over the track segments yields total number of DSBs induced by a single track in a cell nucleus. With access to distribution of DSBs in a single cell, we repeat the above procedures to calculate DSBs induced by collection of independent proton tracks. By further averaging over ensemble of cells, we obtain statistical mean of DSBs and its higher order moments in a colony of cells.
Hence, the output of ReaxFF MD simulation yields the type and magnitude of the DNA damage in nano-scale as a function of energy loss.
By repeating these steps for different locations of the scanning beam of proton, we calculate the proton quality factor in inducing DNA damage and in particular DSBs as a function of depth and energy of proton in tissue.

The main hurdle for applying microscopic models as such is the analysis and processing of gigabyte data extracted from ionizations and excitations in multi tracks of charged particles that is needed for simulation of DNA damage in a single cell.
This is even more challenging if we consider an ensemble of cells to study the effect of radiation to an organ and a clinical object. In this case, the computational limitations in processing of terabyte data makes the microscopic models impractical for any clinical application such as simulation of tumor response and post irradiation tissue toxicity.
Hence we engage an alternative solution and adopt a higher level of spatiotemporal modeling and interface Geant4DNA-ReaxFF results to macroscopic models.

We organize the rest of this paper to present construction of such higher-level modeling and its application in fitting the cell survival data,
however, it is important to emphasize that it is the multi-scaling that constitutes a framework for construction of coarse-grained modeling employed in this study.
We postpone to present more details of this study to our forthcoming publications.

To build a coarse-grained stochastic model of DSB induction by direct or indirect free radical-mediated ionization effects, we consider passage of a single track of a charged particle as shown in Figures \ref{Fig000x} and \ref{Fig00}.
As pointed out, we divide a DNA material in a cell nucleus into nm-size virtual domains such that in each domain there would be a possibility in scoring at least one DSB with probability $p$.
By performing finite number of Geant4DNA-ReaxFF MD simulations and scoring binary numbers for DSB counting, i.e., 0 and 1 associated with zero and greater than zero DSB configurations, we calculate $p$.
Note that as we commented out earlier, $p$ is a function of proton depth and energy in tissue.
For a very large number of samplings over the entire domains of DNA materials, where scoring the statistics of DSBs by a microscopic method such as Geant4DNA-ReaxFF MD is impractical, a Bernoulli / binomial distribution function can be employed to determine the expected number of DSBs and their higher order moments.

For large number of domains with typical particle fluence optimized for therapeutic applications, it is reasonable to assume $p << 1$ hence Poisson distribution, $P_k = (\Delta^k/k!) e^{-\Delta}$ governs distribution of $k=0,1,2,\dots$ DSBs in DNA material with mean value $\Delta = \overline{k}=\sum_{k=0}^\infty k P_k$.
This assumption is consistent with our scoring of DSBs calculated by  Geant4DNA-ReaxFF simulation.

The probability in scoring exactly $\nu$-events resulted from passage of multi-track of particles is also assumed to be Poisson distribution, $P_\nu = (\Theta^\nu/\nu!) e^{-\Theta}$.
Here $\nu=0,1,2,\dots$ counts number of events scored in DNA material and $\Theta=\overline{\nu}=\sum_{\nu=0}^\infty \nu P_\nu$ is the mean number of events scored in collection of cell nuclei.

Combination of intra- and inter-track energy depositions, constitute a compound Poisson distribution and in particular Neyman's distribution of type A~[\onlinecite{Virsik1981:RR,Gudowska2000:APP}] that governs the probability distribution of $n$ DSBs in DNA material induced by collection of independent events
$Q_n (\Delta; \Theta) = \sum_{\nu=0}^\infty P_\nu(\Theta) P_n(\nu\Delta) = e^{-\Theta} \Delta^n/n! \sum_{\nu=0}^\infty (\nu^n/\nu!)(\Theta e^{-\Delta})^\nu$
where $n$ is the total DSBs in DNA induced by all tracks with mean value calculated by the average number of events, $\Theta$, times the average number of DSBs per event, $\Delta$, hence $\overline{n}=\Theta\Delta$.
Probability of finding DNA material with zero DSBs is given by $Q_0 = \exp\{-\Theta(1-e^{-\Delta})\} \approx e^{-\Theta\Delta} = e^{-\overline{n}}$, an expression that resembles the cell SF.

\begin{figure}
\begin{center}
\includegraphics[width=1.0\linewidth]{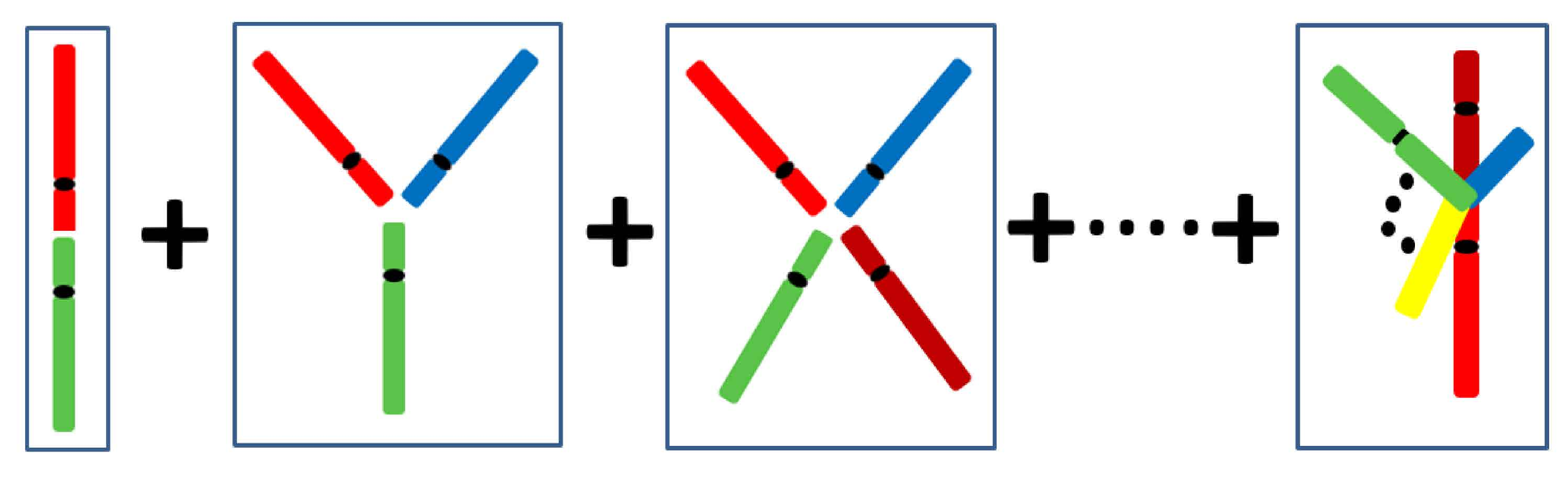}\\ 
\noindent
\caption{
Diagrammatic representation of cell survival in form of a perturbative expansion described by pair-wise, ternary, quaternary, and N-tuple chromosome end-joining corresponding to $\gamma_2$, $\gamma_3$, $\gamma_4$ and $\gamma_N$ in Eq.~(\ref{eq0025Ar}).
Possible combinations of lethal misrepaired lesions are sketched.
}
\label{Fig1_5}
\end{center}\vspace{-0.5cm}
\end{figure}

\subsection{DSB master equation: dynamics}
We now turn to consider the dynamics of DSB's. A time-dependent Master equation, i.e., a birth-death Poisson process, describes the stochastic evolution of DSBs formed on DNAs and post-irradiation intra and inter chromosomes end-joining processes in a cell, including repair and mis-repair mechanisms~[\onlinecite{Albright1989:RR}].
The latter accounts for cellular lethal transitions and/or cell cycle termination.
The transition rates are phenomenological parameters that describe the processes as illustrated in Figures~\ref{Fig00}-\ref{Fig1_5} throughout homologous and non-homologous end joining.
Their numerical values can be extracted by the fitting of the present model to the RBE experimental data~[\onlinecite{Abolfath2017:SR}].

The Poisson process is based on the Markov-chain where events in time depends only on one step behind with no long-range history from the past, as shown schematically in Figure \ref{Fig00}.
The master equation associated with such stochastic process is given by~[\onlinecite{vanKampen:Book}]
\begin{equation}
\frac{dQ_n(t)}{dt} = g_{n-1} Q_{n-1} - g_n Q_n + r_{n+1} Q_{n+1} - r_n Q_n,
\label{eq002}
\end{equation}
where $g_n$, and $r_n$ are creation and annihilation rates respectively corresponding to transitions $n \rightarrow n+1$ and
$n \rightarrow n-1$ in DSB inductions.
$Q_n$ is the normalized ($\sum_{n=0}^\infty Q_n(t)=1$) probability of occurrence of $n$-integer DSBs resulted from scoring of all events in a cell.

It is straightforward to show (see Appendix A) a rate equation that counts for overall damage, including repair and mis-repair mechanisms, can be derived from Eq.~(\ref{eq002})
\begin{eqnarray}
\frac{d\overline{n}(t)}{dt} = \sum_{n=0}^\infty (g_n - r_n) Q_n,
\label{eq0025A}
\end{eqnarray}
where $\overline{n}$ is average number of DSBs in DNA material.
Assuming Neyman's distribution for DSB partition function, $Q_n$, we find the stationary solution of Eq.~(\ref{eq0025A}), $\overline{n}=\Theta\Delta$.
Stationary solution holds only if $r_n=0$ (no repair) post irradiation.

Note that the time evolution of DSBs and their interaction mediated by enzymatic repair mechanisms cannot be simulated by microscopic Geant4DNA-ReaxFF MD, at least at its current stage of development in our computational capability. Hence employing a coarse-grained model, such as the one presented in this subsection is necessary.

\subsection{Chromosome end joining complexities: LET corrections}
\label{sect_2}
The rate of DSB induction can be approximated by $g_n = \mu \dot{z}$ for any $n$. Here $z$ and $\mu$ are the specific energy (deposited energy per mass in unit of Gy) and the average number of DSBs per deposition of 1 Gy of ionizing dose, respectively, and
$\dot{z}$ denotes the specific-energy rate.
The repair and all mis-repair endjoining processes including, unary, binary, ternary, quaternary, ... are incorporated into phenomenological rate constants
\begin{eqnarray}
r_{n} &=& \gamma_1 n + \gamma_2 n(n-1)/2 + \gamma_3 n(n-1)(n-2)/3! \nonumber \\ &&
+ ... + \gamma_N n(n-1)(n-2)...(n-N)/N!.
\label{eq0025Ar}
\end{eqnarray}
The first term in Eq.~(\ref{eq0025Ar}) represents DNA repair processes including DNA non-lethal unary end-joining.
The second term in Eq.~(\ref{eq0025Ar}) represents pair-wise DNA mis-repair processes.
It accounts for combination of two DSBs on any pair of chromosomes out of $n$ initially induced DSBs.
The third term in Eq.~(\ref{eq0025Ar}) represents recombination of three DSBs out of $n$ initially induced DSBs, i.e., ternary DNA misrepair processes.
Similarly the rest of the terms in Eq.~(\ref{eq0025Ar}) describe quaternary, and ... recombination of DSBs as shown in Figure~\ref{Fig1_5}.
We note that in physical processes, fragments of DNA are reattached during the repair processes and not DSBs.
For brevity in our terminology we abbreviated combination of DNA fragments caused by a DSB to ``combination of two or more DSBs''.

Insertion of Eq.~(\ref{eq0025Ar}) in Eq.~(\ref{eq0025A}) and summing $n$ over Neyman distribution function we find
\begin{eqnarray}
\frac{d\overline{n}(t)}{dt} &=& \mu \dot{z} - \gamma_1 \overline{n} - \frac{\gamma_2}{2!}\overline{n(n-1)}
- \frac{\gamma_3}{3!}\overline{n(n-1)(n-2)} \nonumber \\
&-& ... - \frac{\gamma_N}{N!}\overline{n(n-1)(n-2)...(n-N)}.
\label{eq0026xx}
\end{eqnarray}
Expansion of moments of $n$ in Eq.~(\ref{eq0026xx}) can be performed by using the identity
\begin{eqnarray}
\overline{n(n-1)(n-2)...(n-r)}
= \sum_{s=0}^r c_s \overline{n}^{r+1-s} \Delta^s,
\label{eqexp01}
\end{eqnarray}
where $c_s$ are expansion coefficients with $c_0=1$.
Hence Eq.~(\ref{eq0026xx}) can be reduced to a rate equation with rate constants corrected for $\Delta$
\begin{eqnarray}
\frac{d\overline{n}(t)}{dt} = \mu \dot{z} - \lambda_{\rm eff} \overline{n} - \gamma_{\rm eff} \overline{n}^2 - ...,
\label{eq0026}
\end{eqnarray}
Here $\lambda_{\rm eff}$ and $\gamma_{\rm eff}$ are the {\it effective / dressed} chromosome repair and misrepair rates, {\it renormalized} by lineal-energy / LET corrections and can be expressed in power series of $\Delta$, or equivalently LET as we demonstrate below (see e.g., Eq.~(\ref{A29}) and the following discussion).
Hence
\begin{eqnarray}
\lambda_{\rm eff} = \gamma_1 \Lambda(\Delta) = \gamma_1 + \frac{\gamma_2}{2!} \Delta + \frac{\gamma_3}{3!} \Delta^2
+ ... + {\cal O}(\Delta^{N-1}),
\label{eq0027abc}
\end{eqnarray}
and
\begin{eqnarray}
\gamma_{\rm eff} = \frac{\gamma_2}{2!} \Gamma(\Delta) = \frac{\gamma_2}{2!} + \frac{\gamma_3}{3!} 3\Delta + \frac{\gamma_4}{4!} 7\Delta^2
+ ... + {\cal O}(\Delta^{N-2}).
\label{eq0027abs}
\end{eqnarray}
Note that the series in $\gamma_{\rm eff}$ terminates at ${\cal O}(\Delta^{N-2})$ whereas $\lambda_{\rm eff}$ terminates at ${\cal O}(\Delta^{N-1})$.
Here $\gamma_1$ and $\gamma_2/2$ are the {\it bare} chromosome repair and misrepair rates in the absence of corrections from higher order statistical moments where $\Lambda = \Gamma = 1$.
Note that $\Lambda = \Gamma = 1$ implies equivalence of Eq.~(\ref{eq0026xx}) with rate equations considered in previous publications, e.g., in Ref. ~[\onlinecite{Sachs1997:IJRB}] and the references therein where $Q_n$ is assumed to be Poisson distribution and chromosome rejoining mechanisms, beyond binary were disregarded, i.e., $\gamma_3, \gamma_4, ..., \gamma_N$ assumed to be negligible.
The last assumption has been retained in a further extension of Poissonian~[\onlinecite{Sachs1997:IJRB}] to non-Poissonian model of DSB distributions presented in Ref.~[\onlinecite{Carlson2008:RR}].

The relation between dressed and bare parameters resembles renormalization of mass and charge in standard field theories~[\onlinecite{Mandle_Shaw_book}] and effective mass in dispersive media in condensed matter physics~[\onlinecite{Negel_Orlande:book}] with close connection to statistical corrections and zero point fluctuations to the solutions of the Gaussian mean field theories~[\onlinecite{Ramond:book,Kardar:book,Abolfath1998:PRB}].

We denote
\begin{eqnarray}
SF[\overline{n}(t)] = e^{-\overline{L}[\overline{n}(t)]},
\label{eq0027SF}
\end{eqnarray}
the cell survival fraction, and $\overline{L}$ the population of broken DNAs transformed to lethally damaged chromosomes over sufficiently long time, $t\rightarrow\infty$.
$\overline{L}$ can be calculated from fraction of broken DNAs that undergo lethal lesions
\begin{eqnarray}
\overline{L}[\overline{n}(t)] = \int_0^\infty dt \left(\lambda_{L,{\rm eff}} \overline{n}(t) + \gamma_{L,{\rm eff}} \overline{n}^2(t) + ...\right).
\label{eq0027}
\end{eqnarray}
In Eq.~(\ref{eq0027})
\begin{eqnarray}
\lambda_{L,{\rm eff}} = \gamma_{1L} + \frac{\gamma_{2L}}{2!} \Delta + \frac{\gamma_{3L}}{3!} \Delta^2 + ... + {\cal O}(\Delta^{N-1}),
\label{eq0027uvw}
\end{eqnarray}
and
\begin{eqnarray}
\gamma_{L,{\rm eff}} = \frac{\gamma_{2L}}{2!} + \frac{\gamma_{3L}}{3!} 3\Delta + \frac{\gamma_{4L}}{4!} 7\Delta^2 + ... + {\cal O}(\Delta^{N-2}).
\label{eq0027uvz}
\end{eqnarray}
$\gamma_{1L}$, $\gamma_{2L}$, ... $\gamma_{NL}$ represent transition rates of fraction of DSBs that turn to lethal lesions through mis-repair processes of unary, pair-wise, ternary, and higher order DSB recombinations.
Similar to $\lambda_{\rm eff}$ and $\gamma_{\rm eff}$, we can express $\lambda_{L,{\rm eff}}$ and $\gamma_{L,{\rm eff}}$ in power series of particle LET.

\subsection{DSB - Lineal energy dependence}
To calculate the explicit form of $\alpha$ and $\beta$ in linear-quadratic model as a function of lineal-energy / LET and dose, we go one step back and expand Eq.~(\ref{eq0026xx}) in the following form
\begin{eqnarray}
\frac{d\overline{n}(t)}{dt} = \mu \dot{z} - \lambda \overline{n} - \gamma \overline{n^2} - {\cal O}(\overline{n^3}).
\label{eq0026_x}
\end{eqnarray}
Similarly
\begin{eqnarray}
\overline{L} = \int_0^\infty dt \left(\lambda_L \overline{n}(t) + \gamma_L \overline{n^2}(t) + {\cal O}(\overline{n^3})\right).
\label{eq0027_x}
\end{eqnarray}
Eqs.(\ref{eq0026_x}) and (\ref{eq0027_x}) are more general than their MKM~[\onlinecite{Hawkins1998:MP,Hawkins2003:RR}] and RMF models~[\onlinecite{Curtis1986:RR,Sachs1997:IJRB,Carlson2008:RR}] counterparts.
In particular, in MKM and RMF-type models, the parameters $\lambda$, $\gamma$, ${\lambda }_L$ and ${\gamma }_L$ were expanded up to a linear order in LET.
In this work, however, we derive systematically expansion of these parameters to higher orders in LET from Eqs.(\ref{eq0026_x}) and (\ref{eq0027_x}).

To this end, we further split $\overline{n}$ in Eq.~(\ref{eq0026_x}) into $n_0$ and $n_1 = \overline{n} - n_0$.
$n_0$ is a solution of linearized rate equation, i.e., $\gamma=0$ in Eq.~(\ref{eq0026_x}).
A perturbative expansion of $\overline{n}$ around $n_0$ yields $\overline{L}$ in terms of a power series in specific energy fluctuations (see the following section for details)
\begin{eqnarray}
\overline{L} &=& \frac{{\lambda }_L}{\lambda }\mu \overline{z}
+\frac{1}{2}\left[\frac{{\lambda }_L}{\lambda }\frac{\gamma }{\lambda }+\frac{{\gamma }_L}{\lambda }\right]{\mu }^2\overline{z^2}
+\frac{1}{3}\frac{{\gamma }_L}{\lambda }\frac{\gamma }{\lambda }{\mu }^3\overline{z^3}
\nonumber \\ &&
+\left[-\frac{{\gamma }_L}{6\lambda }\frac{{\gamma }^2}{{\lambda }^2}+{\cal O}\left(\frac{{\gamma }^2}{{\lambda }^2}\right)\right] \mu^4\overline{z^4} +{\cal O}\left(\mu^5\overline{z^5}\right).
\label{A13}
\end{eqnarray}

To determine the functionality of $\alpha$ and $\beta$ on LET and LQ cell response, we further expand moments of mean specific energy, $\overline{z^i}$ for any power $i$, in terms of lineal-energy by considering a typical normalized energy deposition distribution function~[\onlinecite{Rossi1996:Book,Kellerer1985:Book}] in a 3 to 7 micrometer size cell nucleus~[\onlinecite{Guan2015:SR}]
\begin{eqnarray}
F\left(z; \Theta\right)=\sum^{\infty }_{\nu =0}{P_{\nu }\left(\Theta\right)}f_{\nu }\left(z\right),
\label{A23}
\end{eqnarray}
Here $f_{\nu}\left(z\right)$ is the distribution of specific energy within $z$ and $z+dz$ in cell nucleus, imparted by exactly $\nu$ energy deposition events from passage of multi-track of particles.
Normalization of $F$ implies $f_\nu$ to be normalized, i.e., $1=\int_0^\infty dz F\left(z; \Theta\right) = \int_0^\infty dz f_\nu\left(z \right)$ for any $\nu$ as $\sum^{\infty }_{\nu =0}{P_{\nu }\left(\Theta\right)} = 1$.

The occurrence of $n = \nu \Delta$ DSBs resulted from energy deposition to the entire DNA material in cell nucleus requires balance in energy transfer, from radiation source to DNA hence
\begin{eqnarray}
f_{\nu }\left(z\right)=\sum^{\infty}_{n=0}{P_n\left(\mu z\right)\delta (z-\nu \Delta/\mu )}.
\label{A29abc}
\end{eqnarray}
Insertion of DNA final state and the DSB density of states in form of $\delta$-function in $f_{\nu}$, enforces a constraint on the energy transfer balance, in accordance with Fermi golden rule~[\onlinecite{SakuraiQM}].
For the consistency checks, readers can easily calculate the norm of $f_{\nu }$ from the last equation and show $1=\int_0^\infty dz f_{\nu }\left(z\right) = \sum^{\infty}_{n=0}P_n\left(\nu \Delta\right)$ for any $\nu$.
Moreover one can show $f_\nu(z) = \int dz' f_1(z') f_{\nu-1}(z-z')$ where $f_1(z)$ is the single event specific energy distribution function.
Substituting $f_{\nu }\left(z\right)$ in Eq.~(\ref{A29abc}) to Eq.~(\ref{A23}) and integrating over $z$ yields both the DSB partition function,
$Q_n\left(\Delta;\Theta\right) = \sum^{\infty }_{\nu =0} P_{\nu}\left(\Theta\right) P_n\left(\nu \Delta\right)$, and correlates $\Delta$ with $z_D$,
the standard deviation of $z$ over whole cell domain population
\begin{eqnarray}
z_D = \frac{\overline{z^2}-\overline{z}^2}{\overline{z}}
=\frac{\Delta}{\mu},
\label{A29}
\end{eqnarray}
where $\overline{z^2} = \int^{\infty }_0 dz {z^2}F\left(z; \Theta\right) = D(D + \Delta/\mu) = D(D+z_D)$ and $\overline{z} = \int^{\infty }_0 dz z F\left(z; \Theta\right) = \Theta\Delta/\mu = \overline{n}/\mu = D$ is the macroscopic / mean dose.


Hence, insertion of $f_\nu$ given in Eq.~(\ref{A29abc}) to Eq.~(\ref{A23}) weighted to powers of $z$ following by integration over $z$ and summing over $\nu$, lead to analytical equations for $\alpha$ and $\beta$ where $\overline{z} = D$, $\overline{z^2} = D(D + z_D)$, $\overline{z^3} = D^3 + 3 z_D D^2 + z^2_D D$, $\overline{z^4} = D^4 + 6 z_D D^3 + 18 z^2_D D + z^3_D D, \cdots$.
Insertion of these identities in Eq.~(\ref{A13}) account for the spatial averaging of the energy deposition fluctuations in cell radiation lethality or toxicity.
In general, calculation of higher order fluctuations in specific energy over the cell nuclei, $\overline{z^i}$, can be recursively expanded to lower power moments, $\overline{z^i}=\sum^i_{j=1}{z^{i-j}_{ij}}{\overline{z}}^j = \sum^i_{j=1}{z^{i-j}_{ij}} D^j$. Here $z^{i-j}_{ij}$ are expansion coefficients derived exactly from Neyman's distribution functions.
$\overline{z^i}$ can be converted to power series in terms of a single variable, $z_D$
\begin{eqnarray}
\overline{z^i}&& = D^i+\left(b_{01}+b_{11}z_D\right)D^{i-1}
\nonumber \\ &&
+ \left(b_{02}+b_{12}z_D\mathrm{+}b_{22}z^2_D\right)D^{i-2} + \dots
\nonumber \\ &&
+ \left(b_{0i}+b_{1i}z_D\mathrm{+}b_{2i}z^2_D+\dots +b_{ii}z^i_D\right)D.
\label{A31}
\end{eqnarray}
$b_{ij}$ are numerical coefficients, independent of $z_D$ and dose with numerical values calculated formally throughout an inverse transformation $z^{i-j}_{ij}=\sum_{k=0}^{i-j} b_{k,i-j} z^k_D$.
In deriving the above equation one can start from
$\overline{(\mu z)^i} = \int_0^\infty dz (\mu z)^i F(z;\Theta) = \sum_{n=0}^\infty n^i Q_n(\Delta; \Theta)$,
which allows use of transformation $\overline{n^i} \rightarrow \overline{(\mu z)^i}$ in Eq.~(\ref{eqexp01}) and turn it to a form given by  Eq.~(\ref{A31}).
As pointed out above, and can be seen in Eq.~(\ref{A31}), $\overline{z^i}$ can be expressed in a power series of a single microdosimetry variable, $z_D$. This is seemingly character of Neyman, Poisson and Gaussian distribution functions as expansion of $\overline{z^i}$ cannot be represented only in terms of powers of $z_D$ for a general distribution function, i.e., the one calculated by MC.

With these identities, one can convert the DSB distribution function, $Q_n$, and express it in terms of deposited dose, $D$, and calculate the cell survival in multi-target theories~[\onlinecite{Vassiliev2012:IJROBP,Vassiliev2017:PMB}]. However, because of lack of postirradiation microscopic biological responses in these theories and adhoc mechanistic assumptions on cell deactivation after specific number of hits as well as lack of clarity in dependence of biological parameters on lineal-energy, we carry on with calculation of the cell-survival throughout the dynamical chromosome repair and mis-repair processes and temporal fluctuations in Markov chain, that yields
\begin{eqnarray}
-{\rm ln (SF)} = \alpha D + \beta D^2 + {\cal O}(D^3),
\end{eqnarray}
where
\begin{eqnarray}
\alpha &=&
\frac{{\lambda }_L}{\lambda}\mu
+ \frac{1}{2}\left[\frac{{\lambda }_L}{\lambda }\frac{\gamma }{\lambda}+\frac{{\gamma }_L}{\lambda }\right]{\mu }^2 z_D
+ \frac{1}{3}\frac{{\gamma }_L}{\lambda }\frac{\gamma }{\lambda }{\mu }^3 z^2_D
\nonumber \\
&+&\left[-\frac{{\gamma }_L}{6\lambda }\frac{{\gamma }^2}{{\lambda }^2} + {\cal O}\left(\frac{{\gamma }^2}{{\lambda }^2} \right)\right] \mu^4 z^3_D + {\cal O}\left(\mu^5 z^4_D\right),
\label{Ney_alpha}
\end{eqnarray}
and
\begin{eqnarray}
\beta &=&
\frac{1}{2}\left[\frac{{\lambda }_L}{\lambda }\frac{\gamma }{\lambda}+\frac{{\gamma }_L}{\lambda }\right]{\mu }^2
+ 3 \frac{1}{3}\frac{{\gamma }_L}{\lambda }\frac{\gamma }{\lambda }{\mu }^3 z_D
\nonumber \\
&+&18 \left[-\frac{{\gamma }_L}{6\lambda }\frac{{\gamma }^2}{{\lambda }^2} + {\cal O}\left(\frac{{\gamma }^2}{{\lambda }^2}  \right)\right] \mu^4 z^2_D + {\cal O}\left(\mu^5 z^3_D\right).
\label{Ney_beta}
\end{eqnarray}

Specific relations and conditions among expansion coefficients in $\alpha$ and $\beta$, as shown in Eqs.~(\ref{Ney_alpha}-\ref{Ney_beta}), are characteristics of Neyman's distribution function.
To employ a similar model for fitting the SF data, it would be reasonable to score the distribution functions directly by performing MC and subsequently fitting the numerically calculated energy loss spectrum to Landau / Vavilov distribution functions~[\onlinecite{Landau1944:JP,Vavilov1957:JETP}] as discussed in Ref.~[\onlinecite{Abolfath2017:SR}].
Because these distribution functions are slightly different and show deviation from Neyman's distribution function, they result in $\alpha$ and $\beta$ in series expansion of moments of $z$, but in more complex form compare to Eqs.~(\ref{Ney_alpha}-\ref{Ney_beta}).
In general
\begin{eqnarray}
\alpha &=&\frac{{\lambda }_L}{\lambda }\mu +\frac{1}{2}\left(\frac{{\lambda }_L}{\lambda }\frac{\gamma }{\lambda }+\frac{{\gamma }_L}{\lambda }\right){\mu }^2z_{21} \nonumber \\
&+&\frac{1}{3}\frac{{\gamma }_L}{\lambda }\frac{\gamma }{\lambda }{\mu }^3 z^2_{31}+\dots,
\label{A31_1}
\end{eqnarray}
and
\begin{eqnarray}
\beta =\frac{1}{2}\left(\frac{{\lambda }_L}{\lambda }\frac{\gamma }{\lambda }+\frac{{\gamma }_L}{\lambda }\right){\mu }^2+\frac{1}{3}\frac{{\gamma }_L}{\lambda }\frac{\gamma }{\lambda }{\mu }^3 z_{32}+\dots.
\label{A31_2}
\end{eqnarray}
Except $z_{21}$ that is identical to $z_D$, the other coefficients, $z_{ij}$, are not necessarily proportional to $z_D$, however they can be expressed as a non-linear function of $z_D$.
We postpone the details in derivation of these equations to Sec. \ref{NLSolutions}.

\subsection{Lineal energy - LET dependence}
\begin{figure}
\begin{center}
\includegraphics[width=1.0\linewidth]{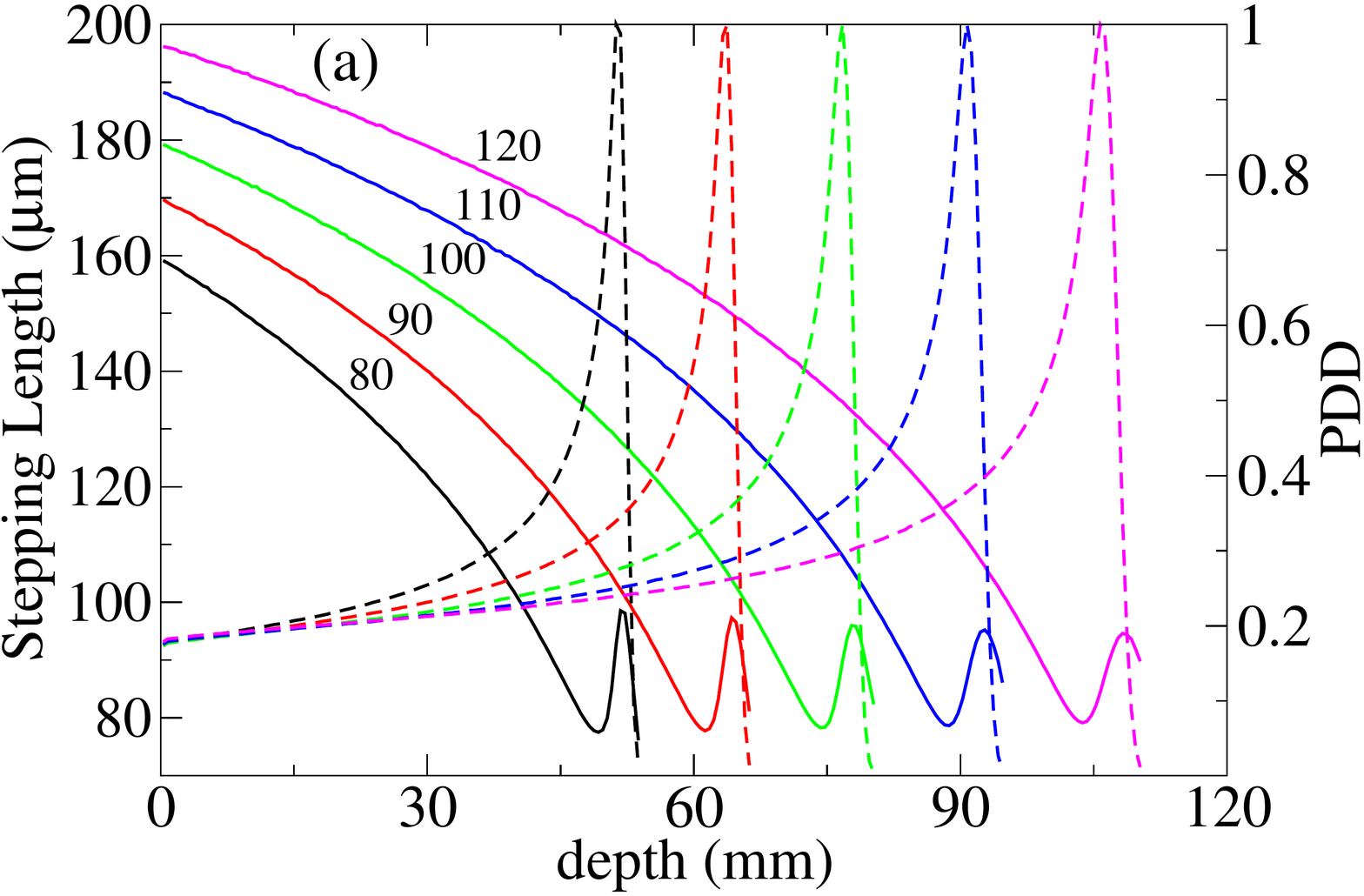}\\ 
\includegraphics[width=1.0\linewidth]{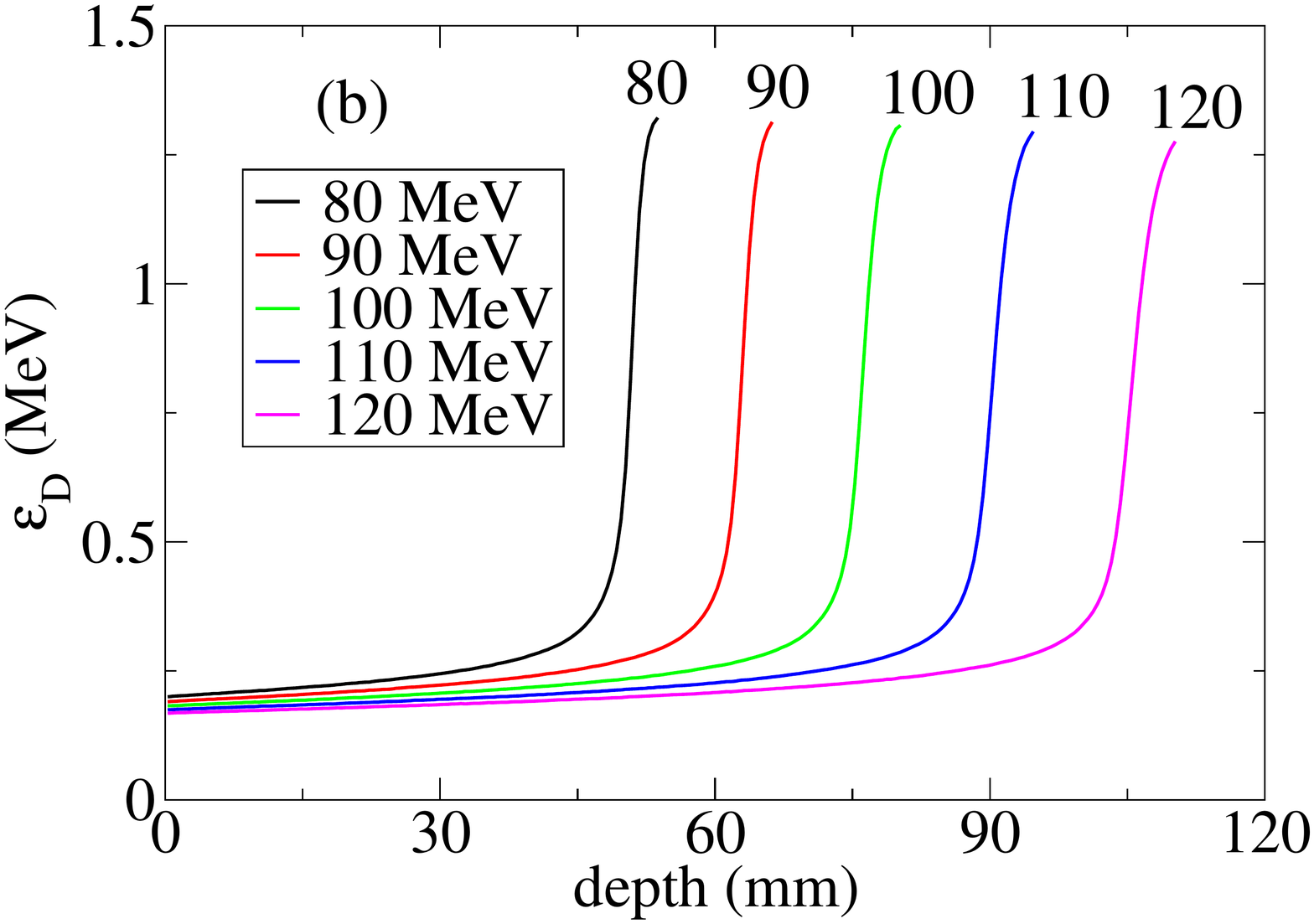}\\ 
\includegraphics[width=1.0\linewidth]{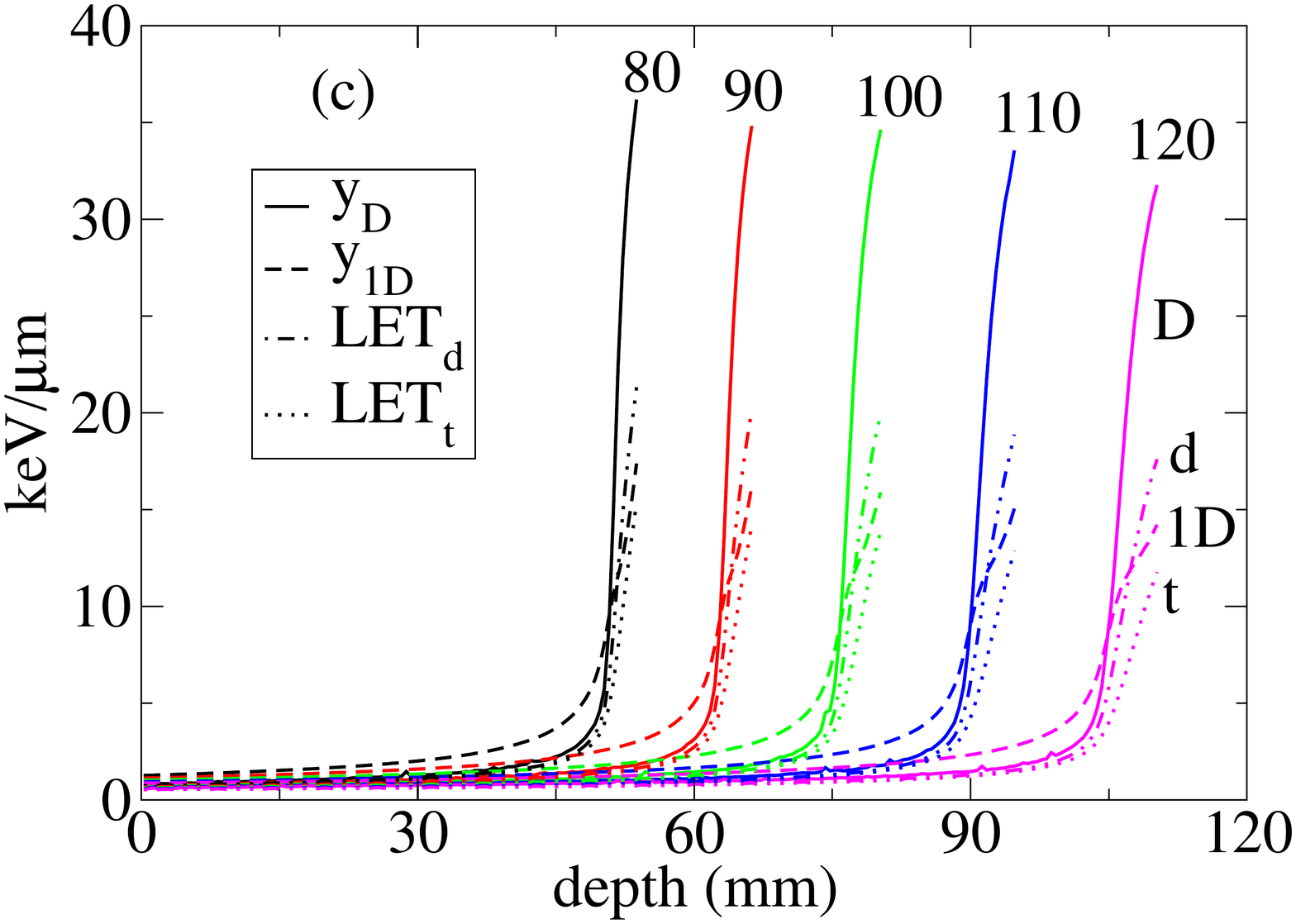}\\ 
\noindent
\caption{
Shown (a) $\overline{l}$ and energy deposition percentage depth dose normalized to its value at the Bragg peak (PDD) (b) energy deposition modeled in SF, $\varepsilon_D = \overline{\varepsilon^2}/\overline{\varepsilon}$ and (c) four types of LET averaging used in this work vs. depth for pencil beams of protons with nominal energies, 80, 90, 100, 110, and 120 MeV.
The letters $D, d, t$ and 1D, denote lineal-energy averaged LET, $y_D=\overline{y^2}/\overline{y}$, dose-averaged LET, LET$_d$, and two types of track-averaged LETs, ${\rm LET}_t = \overline{\varepsilon}/\overline{l}$ and $y_{1D} = \varepsilon_D / \overline{l}$ respectively.
}
\label{fig1xxxxz}
\end{center}\vspace{-0.5cm}
\end{figure}

To apply above formulations to numerical fitting procedure and the experimental data, change of variables and more specifically transformation from $z_D$ to $y_D$ and subsequently to LET$_d$ is required.
To proceed, we define $z_{D}=\overline{l}(y_{1D}/m)$ where $\overline{l}$ is the mean stepping length and $m$ is mass of DNA material under exposure.
In contrast to standard microdosimetry where $\overline{l}$ is defined as the mean chord length, a constant specific to geometrical structure of a microdosimetry volume, we consider $\overline{l}$ a variable that can be calculated by event by event MC simulation of track structures in a voxel in the target volume,
hence $\overline{l}$ is a function of depth as shown in Figure~\ref{fig1xxxxz}, i.e., independent of proton energies.
In Figure~\ref{fig1xxxxz}(a) $\overline{l}$ and energy deposition, normalized to its value at the Bragg peak, e.g., the beam percentage depth dose (PDD), is shown as a function of depth.
In high energies, in the beam entrance, $\overline{l}$ varies as a function of beam energy. In low energies, in the vicinity of the Bragg peak, $\overline{l}$ shows no variation in beam energy, a universal character. In Figure~\ref{fig1xxxxz}(b) the energy deposition in SF, $\varepsilon_D = \overline{\varepsilon^2}/\overline{\varepsilon}$ is shown and in Figure~\ref{fig1xxxxz}(c) four types of LET averaging, $y$-averaged LET, $y_D=\overline{y^2}/\overline{y}$, dose-averaged LET, LET$_d$ and two types of track-averaged LETs, ${\rm LET}_t = \overline{\varepsilon}/\overline{l}$ and  $y_{1D} = \varepsilon_D / \overline{l}$ vs. depth are depicted.

Out of these four LET's, $y_{1D}$, as defined in this study, has been used in formulation of SF where
by replacing $z_D = \Delta/\mu$ we may introduce a quantity that can be interpreted as a measure of number of induced DSBs per track length, $\Delta_l = \Delta / \overline{l}$
\begin{eqnarray}
\frac{\Delta}{\overline{l}} = \frac{\mu}{m} y_{1D}.
\label{A29xxy}
\end{eqnarray}
Similar to $y_{1D}$, $\Delta_l$ is a function of depth and can be calculated by event by event MC simulation as described above.

Calculation of $y_{1D}$ is based on single event specific distribution function where $z_{D} = \varepsilon_D/m = \overline{z_1^2}/\overline{z_1}$.
Here $\overline{z_1} = \int dz z f_1(z) = z_F$ and  $\overline{z_1^2} = \int dz z^2 f_1(z)$, hence $y_{1D} = m (\overline{z_1^2}/\overline{z_1})/\overline{l}$.
The equivalence between $z_D$ in single event and multi-events given by Eq.~(\ref{A29}) can be found through the Fourier transforms of $F(z;\Theta)$ and $f_\nu(z)$ subjected to convolution of $f_\nu$'s, $f_\nu(z) = \int dz' f_1(z') f_{\nu-1}(z-z')$, as described in Ref.~[\onlinecite{Rossi1996:Book}].
More explicitly, $z_{D} = \overline{z_1^2}/\overline{z_1} = (\overline{z^2}-\overline{z}^2)/\overline{z}$.
We apply these equations in average number of DSB per track, $\Delta = \mu z_D$, and subsequently to Eqs.(\ref{eq0027abc}-\ref{eq0027abs}) and Eqs.(\ref{eq0027uvw}-\ref{eq0027uvz}) to express repair-misrepair effective rates
\begin{eqnarray}
\lambda_{\rm eff} &=& \gamma_1 + \frac{\gamma_2}{2!} \mu z_D + \frac{\gamma_3}{3!} \mu^2 z^2_D
+ ...  \nonumber \\
&+& {\cal O}\left((\mu z_D)^{N-1}\right),
\label{eq0027abc1}
\end{eqnarray}
and
\begin{eqnarray}
\gamma_{\rm eff} &=& \frac{\gamma_2}{2!} + \frac{\gamma_3}{3!} 3\mu z_D + \frac{\gamma_4}{4!} 7\mu^2 z^2_D
+ ...  \nonumber \\
&+& {\cal O}\left((\mu z_D)^{N-2}\right),
\label{eq0027abs2}
\end{eqnarray}
and their lethal lesion counterparts
\begin{eqnarray}
\lambda_{L,{\rm eff}} &=& \gamma_{1L} + \frac{\gamma_{2L}}{2!} \mu z_D + \frac{\gamma_{3L}}{3!} \mu^2 z^2_D + ... \nonumber \\
&+& {\cal O}\left((\mu z_D)^{N-1}\right),
\label{eq0027uvw1}
\end{eqnarray}
and
\begin{eqnarray}
\gamma_{L,{\rm eff}} &=& \frac{\gamma_{2L}}{2!} + \frac{\gamma_{3L}}{3!} 3\mu z_D + \frac{\gamma_{4L}}{4!} 7\mu^2 z^2_D + ... \nonumber \\
&+& {\cal O}\left((\mu z_D)^{N-2}\right).
\label{eq0027uvz1}
\end{eqnarray}

It is also straightforward to calculate mean frequency (single event) specific energy, $z_F = \int^{\infty }_0 z f_1(z) = \Delta/\mu$, hence we obtain $z_F = z_D$.
Note that this equation does not hold in general.
It is seemingly a result of particular symmetries and choice of Neyman's distribution function, we considered for our analytical studies.
Similar to multi-event processes, $\alpha$ and $\beta$ can be calculated by Eqs.~(\ref{A31_1}-\ref{A31_2}).
By expanding $z_{ij}$ in Eqs.~(\ref{A31_1}-\ref{A31_2}) around $z_D$, e.g., $\overline{z^n} = \int^{\infty }_0 dz z^n f_1(z) = \int^{\infty }_0 dz (z_D+\delta z)^n f_1(z) = \sum^n_{k=0} \frac{n!}{k! (n-k)!} z^k_D \int^{\infty}_0 dz (z-z_D)^{n-k} f_1(z) = \sum^n_{k=0} \frac{n!}{k!\left(n-k\right)!} z^k_D{\tilde{f}}_{n-k}(z_D)$
where $\delta z=z-z_D$ and ${\tilde{f}}_k\left(z_D\right)=\int^{\infty }_0{{\left(z-z_D\right)}^kf_1(z)dz}$, and using $z_{ij}=\overline{l}y_{ij}/m$,
one may find $\alpha $ and $\beta $ to be a power series in $y_{1D}$, $\alpha = \sum_{i=0}^N a'_i y^i_{1D}$ and $\beta = \sum_{i=0}^{N-1} b'_i y^i_{1D}$.
Here $a'_i$ and $b'_i$ are considered as adjustable phenomenological expansion coefficients.

\begin{figure}
\begin{center}
\includegraphics[width=1.0\linewidth]{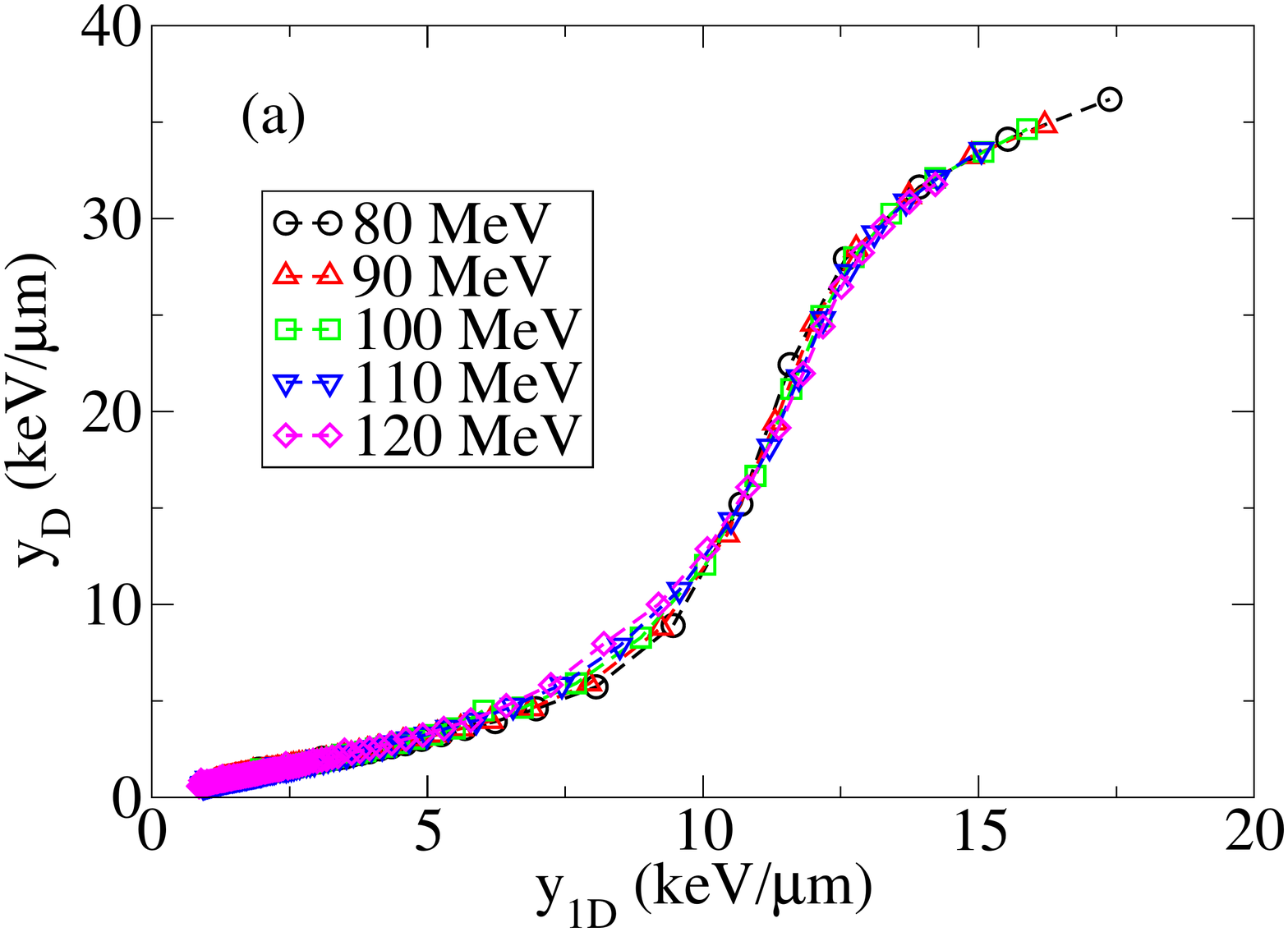}\\ 
\includegraphics[width=1.0\linewidth]{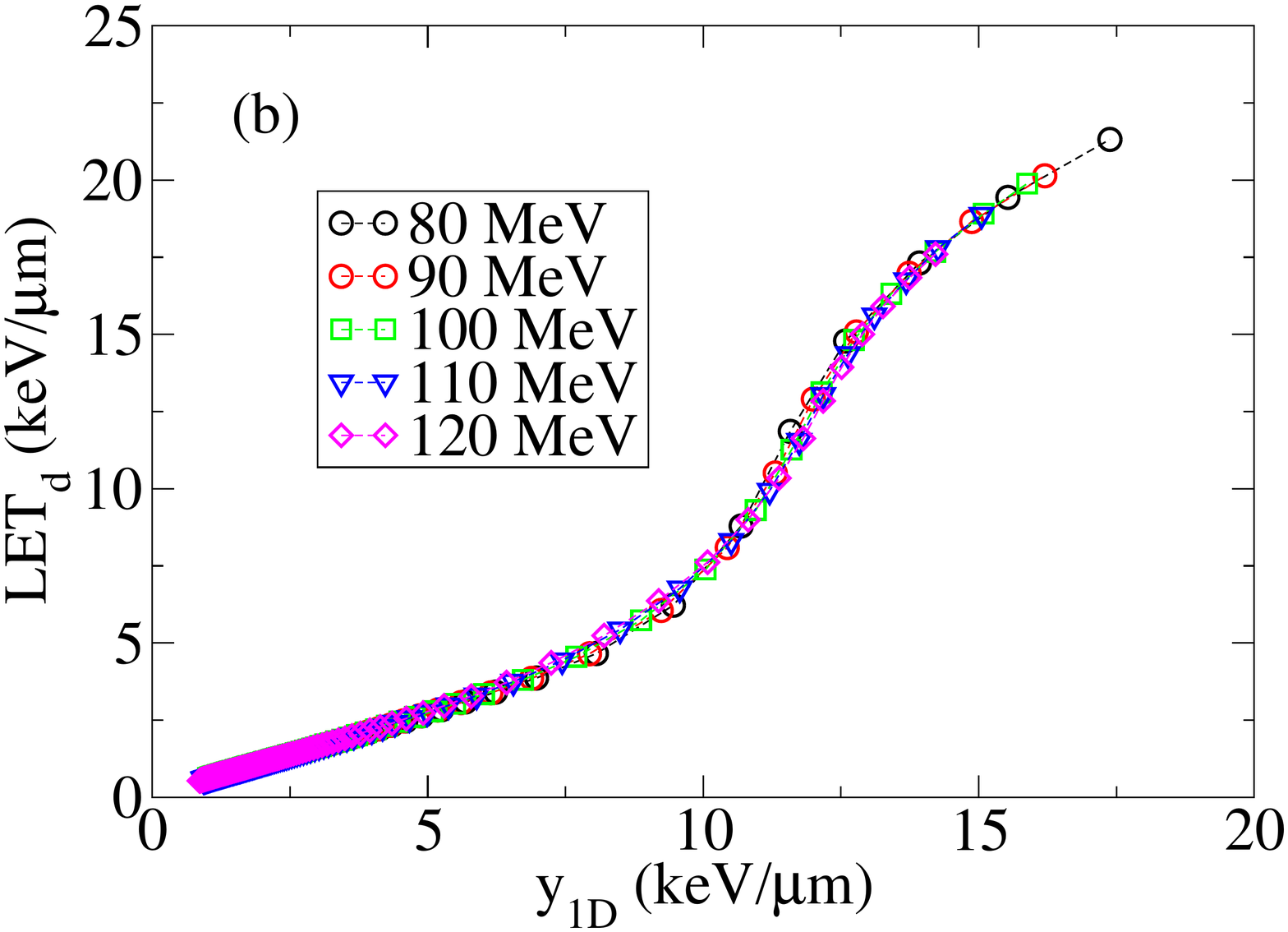}\\ 
\includegraphics[width=1.0\linewidth]{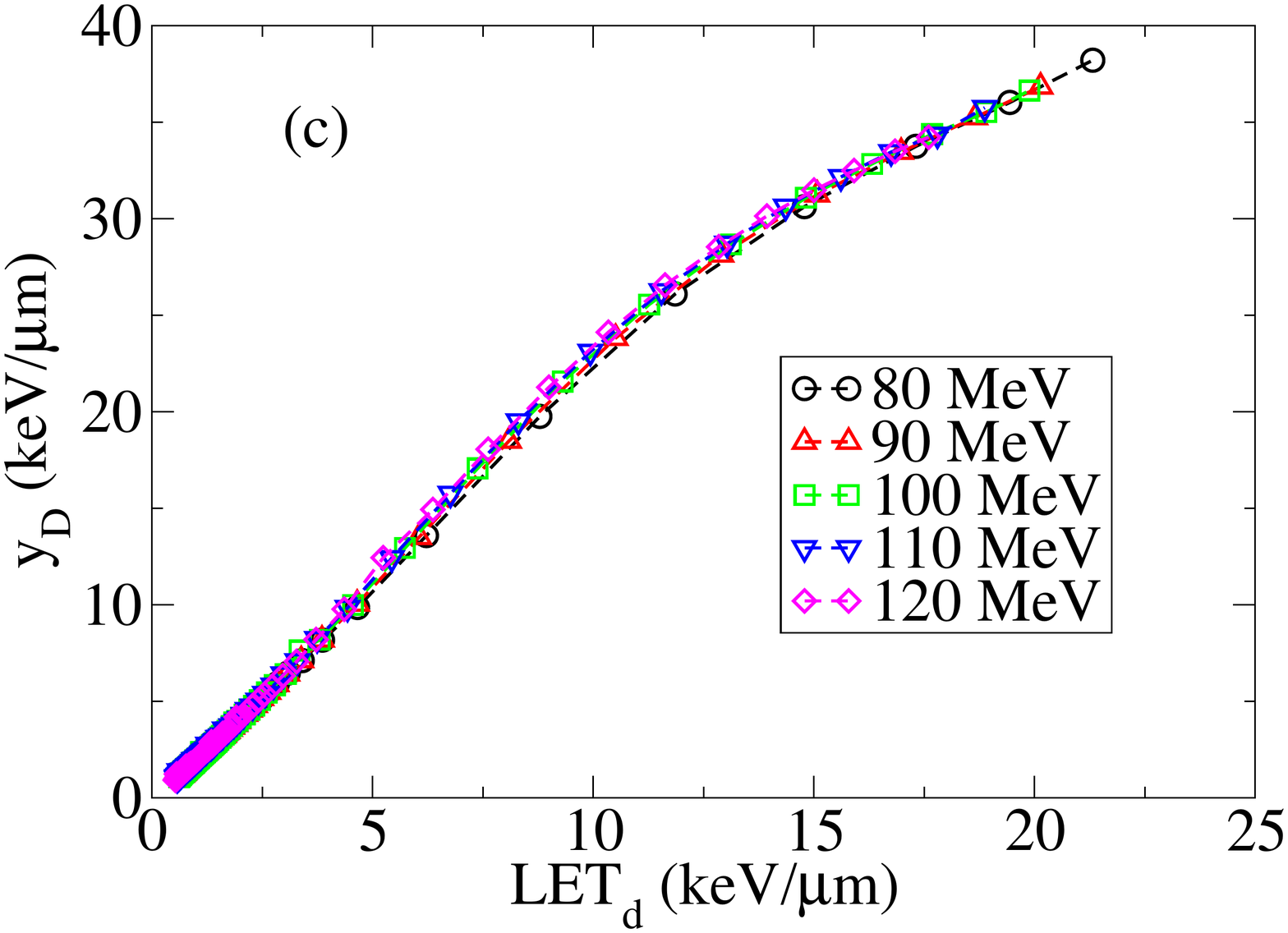}\\ 
\noindent
\caption{
Shown (a) $y_D$ (or LET$_y$) vs. $y_1D$, (b) LET$_d$ vs. $y_1D$ and (c) $y_D$ vs. LET$_d$ calculated by Geant MC toolkit. The simulation consists of $10^6$ particles in water phantom. A universal linear relation in low LET less than 5 $keV/\mu m$ is visible. As LET increases a non-linear dependence emerges and the lines show slightly divergence because of energy loss straggling in the end of proton range.
}
\label{fig1xxxx}
\end{center}\vspace{-0.5cm}
\end{figure}

Note that subscript 1 introduced in $y_{1D}$ intended to distinguish $y_{1D}=m (\overline{z_1^2}/\overline{z_1})/\overline{l}$ from $y_D = \overline{y^2}/\overline{y}$.
The latter is analogous to $y$-averaged LET, LET$_y$ and applied formally in MKM to describe the linear dependence of $\alpha$ on LET, see, e.g., Eq. (26) in Ref.~[\onlinecite{Hawkins1998:MP}].
Figure \ref{fig1xxxx}(a) illustrates the difference between $y_{1D}$ and $y_{D}$ for series of pencil beams of protons calculated by Geant4 MC toolbox~[\onlinecite{Agostinelli2003:NIMA}].

In MC we scored $y_{1D}$ and $y_{D}$ from event by event energy deposition $d\varepsilon_j$ and stepping length $dl_j$, using the following identities
$y_{1D} = (\sum_j (d\varepsilon_j)^2/\sum_j d\varepsilon_j)/(\sum_j dl_j/\sum_j 1_j)$ and $y_D = \sum_j (d\varepsilon_j/dl_j)^2/\sum_j (d\varepsilon_j/dl_j)$ where sum over $j$ includes all energy deposition events from primary and secondary processes in all steps in a specific voxel, hence $\sum_j 1_j$ represents total number of scored energy deposition events.
Our approach in scoring $y_{1D}$ and $y_D$ (as well as other LETs), is based on drawing random variables, $z$ and $y$, from single event distribution functions $f_1(z)$ and $S(y)$, generated on-the-fly by Geant4 MC toolkit.
Interested readers may find more details in Appendix B, however, our approach is equivalent to the scoring method in microdosimetry and the inchoate distribution of energy transfers, introduced by Kellerer in series of publications, see for example Refs.~[\onlinecite{Kellerer_Rossi1972:CTRR,Kellerer1985:Book,Kellerer1975:REB,Grosswendt2005:RPD}].

To investigate variation of $y_{1D}$ and $y_D$, and their dependence on the experimentally reported dose averaged LET, we employ the above methodology and convert $y_{1D}$ and $y_D$ to LET$_d$.
The results of these calculation for pencil beams of $10^6$ protons with nominal energies, 80, 90, 100, 110 and 120 MeV are shown in Figure~\ref{fig1xxxx} where a universal relation, i.e., independent of proton energies, between different forms of LET's is evident.
The one-to-one correspondence between $y_{1D}$ and LET, as seen in Figure~\ref{fig1xxxx}, rationalize the deduction of the following polynomial expansions $\alpha = \sum_{i=0}^N a_i {\rm LET}^i_d$ and $\beta = \sum_{i=0}^{N-1} b_i {\rm LET}^i_d$.
It is worth mentioning three domains in ${\rm LET}_d$, roughly below, within and above 5 and 15 $keV/\mu m$ where the linear relationships between $y_{1D}$ and LET$_d$ (as well as other LET's) change slope.
In particular, in fitting ${\rm LET}_d$ and dose, $D$, as two independent variables to the experimental SF data, we infer three types of dependencies on LET$_d$, within these three domains.
We return to this point in Sec.~\ref{RBE_Results} where we partition SF data points into classes with low, intermediate and high LET$_d$ values respectively.

Note that in the present study, we refer to high and low LETs within the experimental range, relevant to the cell-survival data points as reported in Ref.~[\onlinecite{Guan2015:SR}] where LET $=$ 20 keV/$\mu$m is the maximum experimentally reported value.
Within this range of LETs, we consider LET $=$ 5 and 15 keV/$\mu$m, two limiting values that separate low and high LET domains, in spite of possibilities in scoring even much higher LETs, i.e., above 40 keV/$\mu$m, close to the last $\mu$m's in tail of proton range, where the last cells experience passage of low energy (less than 500 keV) protons.

\subsection{Linear DSB solutions and linear-quadratic (LQ) cell survival}
\label{Sec_linear_ab}
We now turn to calculate solutions of DSB and lethal lesion rate equations.
First we consider a limiting case for $n$ where a linear approximation in Eq.~(\ref{eq0027_x}) can be obtained by neglecting $\gamma$ in the rate equations.
This solution has been used frequently in modeling RBE in literature.
In particular in MKM, e.g., see Eq. (7) in Ref. [\onlinecite{Hawkins1998:MP}].

Insertion of $\gamma =0$ in the rate equation permits access to analytical solution of Eqs.~(\ref{eq0026_x}) and (\ref{eq0027_x}).
Recalling the Green's function method, we can easily convert the differential equation into an integral equation
\begin{eqnarray}
n_0\left(t\right)=\int^{+\infty }_{-\infty }{dt'}G_r\left(t-t'\right)\dot{z}\left(t'\right).
\label{A4}
\end{eqnarray}
Here $n_0$ is the solution of Eq.~(\ref{eq0026_x}) in the linear approximation. It is straightforward to justify that the homogeneous solution to Eq.~(\ref{eq0026_x}) is identical to zero, thus we do not consider it in Eq.~(\ref{A4}). It is also a straightforward calculation to show the retarded Green's function follows $G_r\left(t-t'\right)=\mu e^{-\lambda \left(t-t'\right)}\theta \left(t-t'\right)$, where $\theta \left(t-t'\right)$ is the Heavyside function, i.e., $\theta =1$ if $t\ge t'$ and 0 otherwise. The steps in calculating $G_r$ include converting the integral equation, Eq.~(\ref{A4}), to a differential equation for $G_r$ by substituting ~(\ref{A4}) in ~(\ref{eq0026_x}) and imposing the initial condition ${\overline{n}}_0=0$ for $t<0$ where $\dot{z}=0$. Similarly we define$\ {SF}_0=e^{-{\overline{L}}_0}$ where ${\overline{L}}_0=\int^{+\infty }_{-\infty }{dt'}\left[{\lambda }_L{\overline{n}}_0+{\gamma }_L\overline{n^2_0}\right]$. Here the bar over $L_0$ denotes energy deposition averaging on the ensemble of cell nuclei domains, specific to a lineal-energy distribution.

For an acute radiation dose, $\dot{z}(t)=z\delta (t)$, the solution of Eq.~(\ref{A4}), $n_0(t)=$ $\mu {ze}^{-\lambda t}\theta \left(t\right)$, leads to ${\overline{L}}_0=\frac{{\lambda }_L}{\lambda }\mu \overline{z}+\frac{{\gamma }_L}{2\lambda }{\mu }^2\overline{z^2}$ where by averaging over the lineal-energy distribution and all cell nuclei and their domains we obtain a linear-quadratic model in cell-survival
\begin{eqnarray}
-{\mathrm{ln} \left(SF\right)\ }=\alpha \overline{z}+\beta {\overline{z}}^2,
\label{A5}
\end{eqnarray}
where $\alpha =\frac{{\lambda }_L}{\lambda }\mu +\frac{{\gamma }_L}{2\lambda }{\mu }^2z_D$ and $\beta =\frac{{\gamma }_L}{2\lambda }{\mu }^2$. Here we use the identity $\overline{z^2}=\overline{z}\left(\overline{z}+z_D\right)$.

We further recall relation $z_D=\overline{l} y_{1D}/m$ where $m=\rho V$ is the average mass in a spherical volume $V$ surrounding the point of energy deposition, as introduced in the inchoate distribution of energy transfers in microdosimetry by Kellerer, Rossi and their colleagues ~[\onlinecite{Kellerer_Rossi1972:CTRR,Kellerer1985:Book,Kellerer1975:REB}] and
$\rho$ is the water equivalent mass density.
Therefore, we obtain
$\alpha =\frac{\lambda_L}{\lambda}\mu +\frac{\gamma_L}{2\lambda}\mu^2\frac{1}{\rho \left(V/\overline{l}\right)}y_{1D}$.
Considering $\alpha$ and $\beta$ as phenomenological parameters and $V/\overline{l} = \pi r^2_d$ we end up with two relations
\begin{eqnarray}
\alpha ={\alpha}_0+\beta \frac{1}{\rho \pi r^2_d}y_{1D},~~ \beta =\beta_x.
\label{A6}
\end{eqnarray}
Note that Eq.~(\ref{A6}) resembles the frequently used relation in literature (see e.g., Eq.(II.28) in Ref.~[\onlinecite{Hawkins1998:MP}] or Eq. (8) in Ref.~[\onlinecite{Kase2006}]) if we assume $z_D=\ell y_D/m$ and $\ell$ denotes the average chord length of a MKM domain.

Considering a piece-wise linear relation between $y_{1D}$ and ${\rm LET}_d$, for a domain of ${\rm LET}_d$, less than $5 keV/\mu m$ and greater than $15 keV/\mu m$, as shown in Figure~\ref{fig1xxxx}(b), one may suggest an approximate linear relation between $\alpha$ and ${\rm LET}_d$, i.e., $\alpha = \alpha_0 + \alpha_1 {\rm LET}_d$.
Here $\alpha_0$, $\alpha_1$, $\beta_0$ and $r_d$ are four fitting parameters with only three independent parameters.
Note that $\alpha_1$ is linearly proportional to $\beta/(\rho \pi r^2_d)$.
Because of piece-wise linear dependence between $y_{1D}$ and LET$_d$, as illustrated in Figure~\ref{fig1xxxx}, the corresponding coefficient changes value at ${\rm LET}_d = 5 keV/\mu m$ and ${\rm LET}_d = 15 keV/\mu m$.

Alternatively one can fit a polynomial to $y_{1D}$-${\rm LET}_d$ curve and obtain $\alpha$ a non-linear function of ${\rm LET}_d$.
Note that, because in Eq.~(\ref{A6}), $\beta$ does not change by variations in LET$_d$ as it is constant and independent of $y_{1D}$, $\alpha / \beta$ increases monotonically as a function of ${\rm LET}_d$.
In the following section, we discuss the presence of other sources that contribute to the non-linearities of $\alpha$ as a function of $y_{1D}$.
Same mechanisms contribute to dependence of $\beta$ on $y_{1D}$.
Therefore we argue that both $\alpha$ and $\beta$ are non-linear functions of LET$_d$.

\subsection{Non-linear expansion of DSB solutions, going beyond LQ cell survival}
\label{NLSolutions}
In this section, we continue with performing a perturbative expansion to calculate the non-linear solution of Eqs.~(\ref{eq0026_x}) and (\ref{eq0027_x}). To go beyond the linear solutions presented in the preceding section, we assume $\gamma $ to be a small parameter, hence we expand $\overline{n}$ about $n_0$ perturbatively and linearize the resulting rate equation to obtain the dynamics of the small fluctuations describing deviations from linear DSB solutions. We define $n_1=\overline{n}-n_0$ and recall Eq.~(\ref{eq0026_x}) to obtain a linear equation for $n_1$
\begin{eqnarray}
\frac{dn_1}{dt}=-\lambda n_1-\gamma \left(2n_0n_1+n^2_0\right)+{\cal O}\left(n^2_1\right).
\label{A5_1}
\end{eqnarray}
Here $\overline{n}$ is the exact solution of Eqs. ~(\ref{eq0026_x}) and ~(\ref{eq0027_x}).
By definition, $n_0$ is the exact linear solution of these equations, hence $n_1$ describes the difference between exact and linear solutions. It is more convenient to transform Eq.~(\ref{A5_1}) into a more compact form
\begin{eqnarray}
\frac{dn_1}{dt}+\eta \left(t\right)n_1\left(t\right)=\xi \left(t\right),
\label{A6_1}
\end{eqnarray}
where $\eta =\lambda +2\gamma n_0$ and $\xi =\gamma n^2_0$. The solution of Eq.~(\ref{A6_1}) can be calculated exactly
\begin{eqnarray}
n_1\left(t\right)=e^{-\varphi (t)}\int^t_{-\infty }{dt' \xi \left(t'\right)e^{\varphi \left(t'\right)},}
\label{A7_1}
\end{eqnarray}
where $\varphi \left(t\right)=\lambda t+2\gamma \int^t_{-\infty }{dt'n_0(t')}$. Linearizing Eq.~(\ref{A7_1}) in terms of $\gamma $, assuming $\gamma $ is a small parameter, leads to
\begin{eqnarray}
n_1\left(t\right)=\gamma e^{-\lambda t}\int^t_{-\infty }{dt'n^2_0(t')e^{\lambda t'}}+{\cal O}\left(\frac{{\gamma }^2}{{\lambda }^2}\right).
\label{A8_1}
\end{eqnarray}

Substituting the linear solution calculated above, $n_0(t)=$ $\mu {ze}^{-\lambda t}\theta \left(t\right)$, in Eq.~(\ref{A8_1}) yields
\noindent
\begin{eqnarray}
n_1\left(t\right)=\frac{\gamma }{\lambda }{\mu }^2z^2\left(1-e^{-\lambda t}\right)e^{-\lambda t},
\label{A9_1}
\end{eqnarray}
hence
\begin{eqnarray}
{{\overline{n}=n}_0+n}_1=n_0-\frac{\gamma }{\lambda }\left[n_0-\mu z\right]n_0+{\cal O}\left(n^3_0\right).
\label{A10_1}
\end{eqnarray}

From Eq.~(\ref{A10_1}) and $n_0$, the cell-survival can be calculated, $-{\mathrm{ln} \left(\mathrm{SF}\right)\mathrm{=}\ }\int^{+\infty }_{-\infty }{dt'}\left[{\lambda }_L\overline{n}+{\gamma }_L\overline{n^2}\right]$, hence we obtain Eq.~(\ref{A13})
as given below
\begin{eqnarray}
-{\rm ln (SF)} &=& \frac{{\lambda }_L}{\lambda}\mu \overline{z} + \frac{1}{2}\left[\frac{{\lambda }_L}{\lambda }\frac{\gamma }{\lambda}+\frac{{\gamma }_L}{\lambda }\right]{\mu }^2\overline{z^2}
+\frac{1}{3}\frac{{\gamma }_L}{\lambda }\frac{\gamma }{\lambda }{\mu }^3\overline{z^3}
\nonumber \\
&+&\left[-\frac{{\gamma }_L}{6\lambda }\frac{{\gamma }^2}{{\lambda }^2}+{\cal O}\left(\frac{{\gamma }^2}{{\lambda }^2}\right)\right]\mu^4\overline{z^4} + {\cal O}\left(\mu^5\overline{z^5}\right).
\label{A11_1}
\end{eqnarray}
The last two terms in Eq.~(\ref{A11_1}) are the contribution of the terms omitted in Eq.~(\ref{A8_1}) due to linearizing $\overline{n}$ in the limit where $\gamma $ is negligible. To transform (\ref{A11_1}) to a form similar to the linear-quadratic model, we must calculate the statistical fluctuations in microscopic dose deposition throughout the averaging over cell nucleus domains, assuming equivalence between the ensemble averaging over the domains and the spatial averaging of the energy deposition fluctuations over the cell nuclei.

Similar to linear model discussed in preceding section, for numerical fitting of the experimental data we transform $y_{1D}$ to LET$_d$ using the relations obtained from MC simulation, as depicted in Figures~(\ref{fig1xxxx}).
However, because in non-linear SF model, both $\alpha$ and $\beta$ vary as a function of $y_{1D}$, with increase of LET beyond 5 $keV/\mu m$, a more rapid non-linearity in $\alpha$, $\beta$ and RBE as a function of LET$_d$ is expected.

Note that the general formalism presented in this study allows insertion of an arbitrary dose rate, $\dot{z}(t)$, in the rate equations where the linear solutions to the cell lethality can be obtained through numerical integration over the dose rate, $\dot{z}(t)$, convolved with the retarded Green's function, $G_r\left(t-t'\right)$ in Eq.~(\ref{A4}).
Subsequently, the numerical integration over the linear solution permits calculation of the non-linear corrections.

\section{Cells birth-death master equation: tumor growth dynamics}
We now turn to incorporate the time evolution of DSBs to the growth rate of the cells, in particular in tumors where
the cells and their offsprings tend to double their population exponentially through mitotic / meiotic cycles.
The growth in cell population is, however, in balance with apoptotic / necrotic cell death pathways.

Although, it is known for decades that the major therapeutic gains in applying ionizing radiation, and in particular, induction of DSBs, is to alter the growth rate balance toward the cell death and disruption of the growth factors and enforcing the tumor shrinkage, but because of deficiencies in quantitative biological optimizations in TPSs, there is still a need for improving model calculations.
There are, however, fundamental computational challenges that does not permit incorporation of the initial events, e.g., the DSB inductions and their interference with the cell growth pathways in the current TPSs.
This is mainly because of the entanglements of the disparities in two vastly separated spatial and temporal landscapes in any computational model as pointed out in the beginning of Sec. ~\ref{sect_1}.
More specifically, while the occurrence of the initial ionizing radiation events and DSB formations have been realized in nano-meter and femto to picosecond scales, the therapeutic / clinical endpoints of the tumor growth can be captured in centimeter and days or even weeks and months.
Thus there are several orders of magnitude difference between initial and final events of interest.
To be able to model the hierarchy of the events and their pathways in a logical and algorithmic methodology, e.g., under a unified simulation toolbox, a coarse-grained multi-scale approach must be engaged to close the gap between microscopic events at the molecular levels and their macroscopic resemblance at the biological / clinical endpoints.
We note that our multi-scale methodology is consistent with a multiscale framework that have been developed and applied successfully to describe cellular response and RBE for ion irradiation by proton and ion beam therapy,~[\onlinecite{Surdutovich2010:PRE,Solovyov2017:Book}].

To achieve this goal, we consider a minimal mathematical approach relevant for cell growth by applying a computational model similar to the time-dependent master equation, proposed in preceding sections to simulate the dynamical processes of DSBs in nano-meter scale.
This model is, however, in macroscopic scale.
To this end, we consider a birth-death Poissonian / Markovian process to describe the stochastic growth of a colony comprise of collection of weakly-interacting cells.
The master equation associated with such processes provides normalized probabilities, $T_N(t)$, with $\sum_{N=0}^\infty T_N(t) = 1$, and $N$ denoted as the number of cells at moment $t$, in a colony that resembles population of $N$ weakly correlated cells
\begin{equation}
\frac{dT_N(t)}{dt} = b_{N-1} T_{N-1} - b_N T_N + d_{N+1} T_{N+1} - d_N T_N.
\label{eq002cds}
\end{equation}
$b_N$ and $d_N$ represent the rates of birth and death in mitotic / meiotic cell division processes and
apoptotic / necrotic mechanisms respectively.
In this model, transitions in cell population only by a single step, $N \rightarrow N+1$, or, $N \rightarrow N-1$, have been taken into account.

It is straightforward to show the rate equation associated with Eq.~(\ref{eq002cds}) is given by (the derivation of this equation is similar to rate equation derived for DSBs and presented in Appendix A)
\begin{eqnarray}
\frac{d\overline{N}(t)}{dt} = \sum_{N=0}^\infty (b_N - d_N) T_N(t).
\label{eq0025Acds}
\end{eqnarray}
For simplicity in modeling, it is customary to consider constant rates, linearly proportional to the cell population, $b_N = b N$ and $d_N = d N$ in the absence of any therapeutic modality such as radiation (see for example, Ref.~[\onlinecite{ZaiderPMB:2000}]), hence $d\overline{N}/dt = (b - d) \overline{N}$.

In the next step, we incorporate the effect of ionizing radiation to Eq.~(\ref{eq0025Acds}) by recalling in-vitro cell survivals, $SF = e^{-\overline{L}}$, calculated in the preceding sections and its basic definition, $SF = N_\infty/N_0$, the ratio of the cells at the biological endpoint, $N_\infty$, survived out of initial number of the seeded cells, denoted by $N_0$.
Therefore a similar rate equation to Eq.~(\ref{eq0025Acds}), can be proposed to correlate $N_\infty/N_0$ with $\overline{L}$ using $SF$ formulation
\begin{eqnarray}
\frac{d\overline{N}(t)}{dt} = - L(t) \overline{N}(t),
\label{eq0026Acds}
\end{eqnarray}
where $L(t)$ is the integrand of $\overline{L}$, i.e., $\overline{L} = \int_0^\infty dt L(t)$.
The time dependence of $L(t)$ stems from the dynamics of DSBs in cell nuclei in microscopic scale where
by referring to Eq.~(\ref{eq0027}), we consider $L(t) = \lambda_{L,{\rm eff}} \overline{n}(t) + \gamma_{L,{\rm eff}} \overline{n}^2(t) + ...$, and from Eq.~(\ref{A10_1}), we have $\overline{n}(t) = n_0(t) - (\gamma/\lambda) (n_0(t)-\mu z) n_0(t) + {\cal O}\left(n^3_0\right)$, and $n_0(t) = \mu z e^{-\lambda t} \theta(t)$.

We further carry on with combining the rate equations, Eqs.~(\ref{eq0025Acds}) and (\ref{eq0026Acds}), that assert $b_N = b N$ and $d_N = d N + L(t)$.
Hence the tumor rate equation, including the ionizing radiation effect, is given by
\begin{eqnarray}
\frac{d\overline{N}(t)}{dt} = \left(b - d - L(t)\right) \overline{N}(t).
\label{eq0026Acds2}
\end{eqnarray}
Similarly the time evolution of $T_N$ is given by
\begin{eqnarray}
\frac{dT_N(t)}{dt} &=& b (N-1) T_{N-1} + \left(d + L(t)\right) (N+1) T_{N+1} \nonumber \\
&-& (b + d + L(t)) N T_N.
\label{eq002cds2}
\end{eqnarray}
From Eq.~(\ref{eq002cds2}), we calculate TCP, under the condition that all irradiated cells were gone through the lethal lesions, and subsequently no tumor cell survived the exposure of ionizing radiation, hence $TCP = T_{N=0}(t=\infty)$.
Calculation of $T_{N}$ using the generating function technique and method of characteristics can be found in Refs.~[\onlinecite{vanKampen:Book,Reichl:Book}] thus we skip to present the details of this calculation.
Accordingly,
\begin{eqnarray}
TCP\left[\overline{n}(t)\right] = \left(1 - SF_{\rm eff} \left[\overline{n}(t)\right]\right)^{N_0},
\label{eq002cds3}
\end{eqnarray}
where
\begin{eqnarray}
SF_{\rm eff}\left[\overline{n}(t)\right]
= \frac{e^{(b - d)t-\overline{L}\left[\overline{n}(t)\right]}}{1 + b e^{(b - d)t-\overline{L}\left[\overline{n}(t)\right]}
\int_{0}^{t} dt' e^{-(b - d)t' + \overline{L}\left[\overline{n}(t')\right]}},
\label{eq002cds3f}
\end{eqnarray}
is effective SF of each individual cell.
From Eq.~(\ref{eq002cds3f}), it should be clear that $\overline{n}(t)$ is a source that resembles driving force in TCP.
Like the action in Lagrangian dynamics of classical particles that is a functional of the particle trajectory~[\onlinecite{Landau_Lifshitz_Mechanics}], $SF_{\rm eff}$ is a {\it functional} of DSB dynamical trajectories $\overline{n}(t)$ where
\begin{eqnarray}
\overline{L}\left[\overline{n}(t)\right] = \int_0^t dt' \left(\lambda_{L,{\rm eff}} \overline{n}(t') + \gamma_{L,{\rm eff}} \overline{n}^2(t') + ...\right),
\label{eq0027zz}
\end{eqnarray}
is a functional integral of $\overline{n}(t)$.
An optimal solution to TCP and SF can be sought by performing a functional variation of $SF_{\rm eff}$ with respect to $\overline{n}(t)$.
A solution to TCP is required to minimize $SF_{\rm eff}$ in the phase space of DSB trajectories
\begin{eqnarray}
\frac{\delta}{\delta\overline{n}(t)} SF_{\rm eff}\left[\overline{n}(t)\right] = 0,
\label{eq002cds3fq}
\end{eqnarray}
where $\delta\overline{n}(t)$ stands for functional derivatives~[\onlinecite{Negel_Orlande:book,Ramond:book,Kardar:book}].
A solution to Eq.~(\ref{eq002cds3fq}) that maximizes $SF_{\rm eff}$ can be considered for normal tissues.

A simple solution to this equation can be derived by neglecting non-locality in time evolution of lethal lesions in Eq.~(\ref{eq002cds3f})
\begin{eqnarray}
\int_{0}^{t} dt' e^{-(b - d)t'} e^{\overline{L}\left[\overline{n}(t')\right]} 
\rightarrow  \frac{1 - e^{-(b - d)t}}{b - d} e^{\overline{L}\left[\overline{n}(t)\right]}.
\label{eq002cds3ffq}
\end{eqnarray}
For an acute dose of radiation in a single fraction at the cell biological endpoint, where $t\rightarrow \infty$, we consider a stationary solution for $\overline{L}$,  a variable that is a constant of time.
In Secs. \ref{Sec_linear_ab} and \ref{NLSolutions}, we showed that $\overline{L}$ follows a linear-quadratic type-relation in dose with $\alpha$ and $\beta$ that are  polynomials of lineal-energy, $y$, and / or LET, i.e., $\overline{L} = \alpha D + \beta D^2 + \gamma D^3 + {\cal O}(D^4)$.
Considering this stationary solution for $\overline{L}$, we subsequently find $SF_{\rm eff} = e^{(b-d)t-\overline{L}}/\left[1 + b/(b-d) (e^{(b-d)t} - 1)\right] \approx e^{-\overline{L}} (b-d)/b$,
as $t \rightarrow \infty$.
For typical cancerous cells, the rate in the cell natural death process is negligible relative to their growth rate, hence $d << b$.
In this limit, Eq.~(\ref{eq002cds3}) can be simplified to
\begin{eqnarray}
TCP(\infty) = \left(1 - e^{- \overline{L}\left[\overline{n}(\infty)\right]}\right)^{N_0}
= \left(1 - SF\right)^{N_0}.
\label{eq002cds3x}
\end{eqnarray}
These equations exhibit interplay of sequence of events starting from DSB induction by ionizing radiation in microscopic scale that propagate to formation of DNA damage and chromosome misrepairs to a coarse-grained tumor dynamical responses in macroscopic scale.
They govern underlying mechanisms that constitute multi-scale formulation of the cell deactivation theory in hadron therapy.


\section{Results}
\subsection{Numerical fitting procedure}
\label{RBE_Results}
\begin{figure}
\begin{center}
\includegraphics[width=0.8\linewidth]{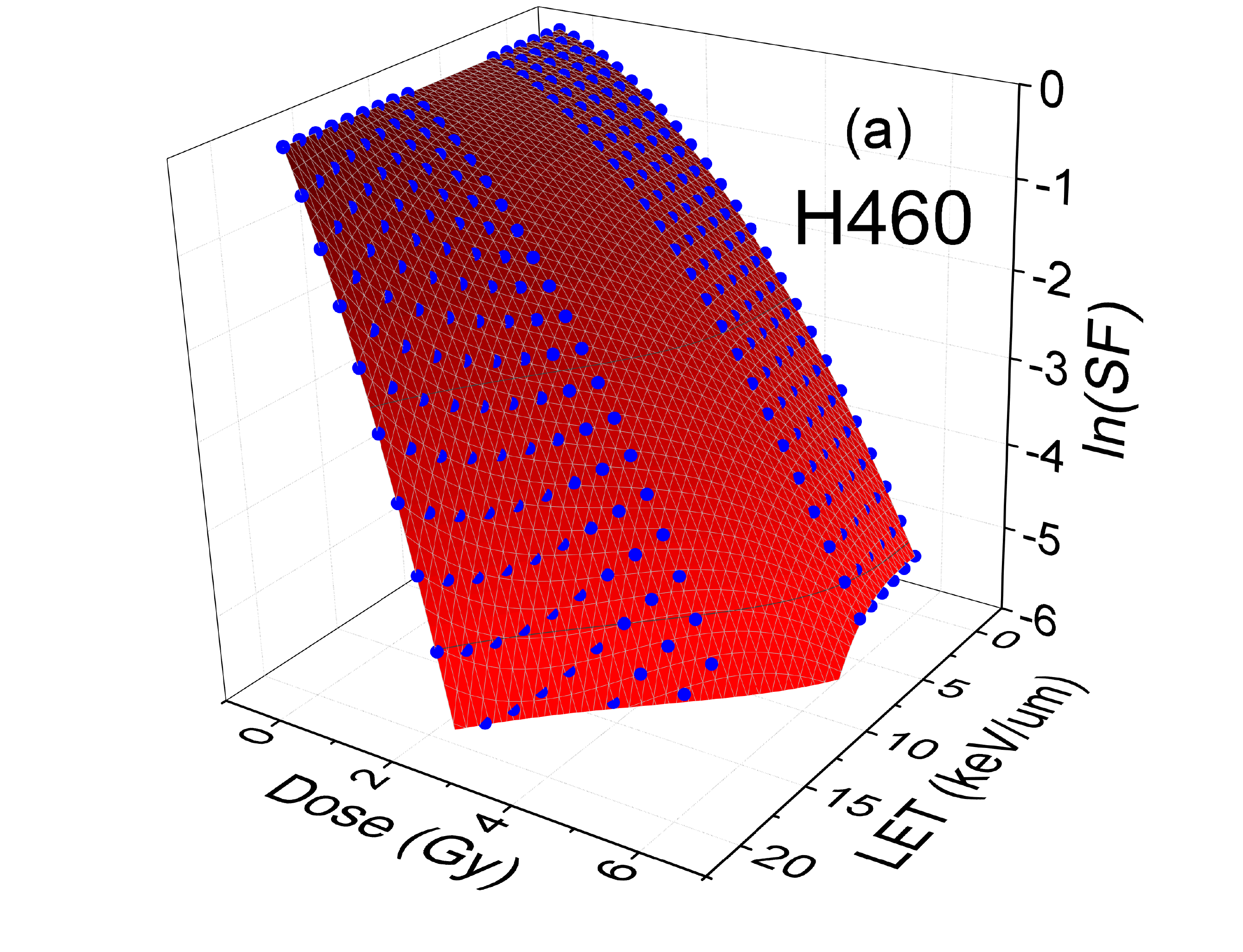}\\ 
\includegraphics[width=0.8\linewidth]{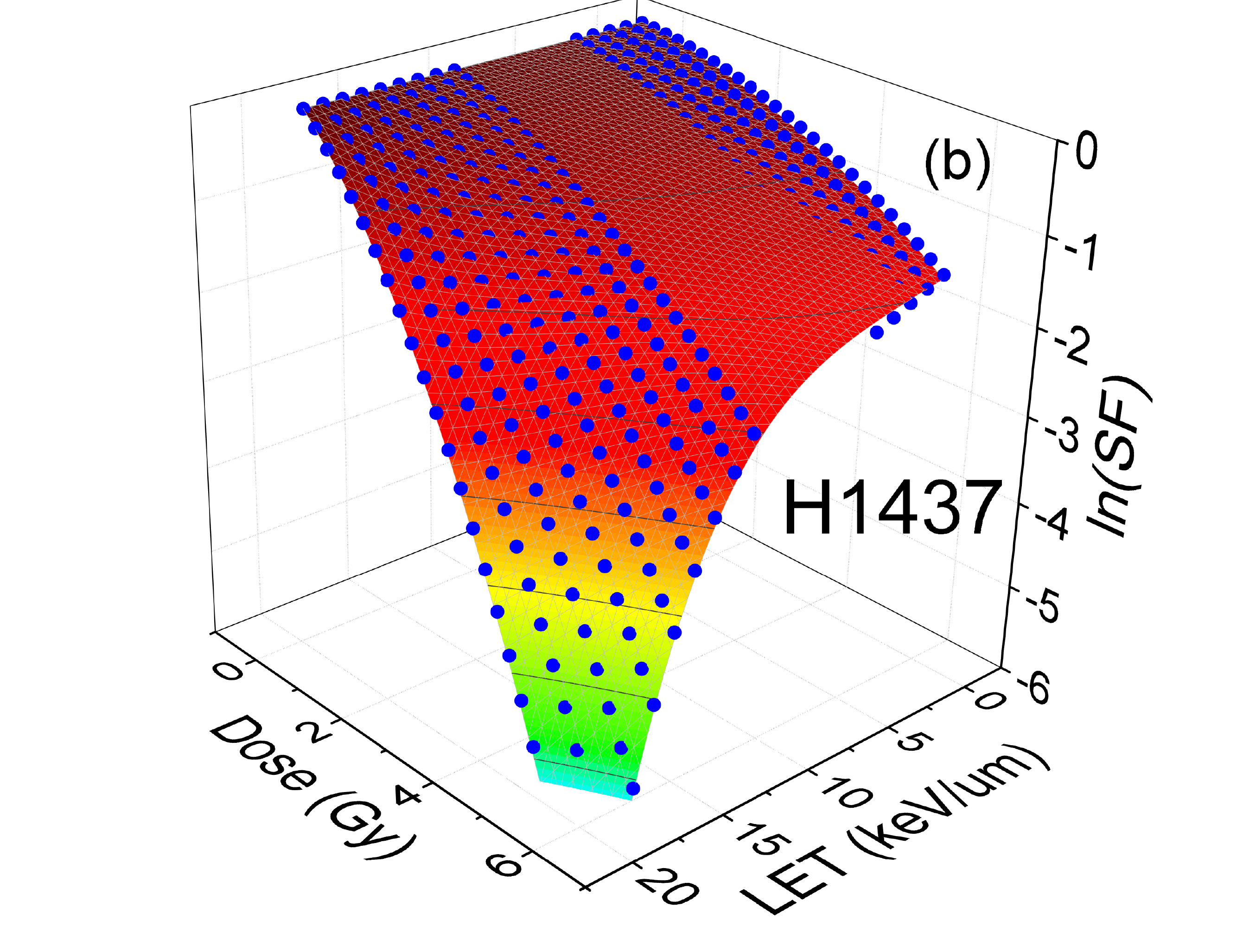}\\ 
\noindent
\caption{
Shown H460 and H1437 cell survival as a function of dose (Gy) and dose averaged LET$_d$ ($keV/\mu m$).
The blue dots are the result of 3D surface fitting to the experimental data presented in Ref.~[\onlinecite{Abolfath2017:SR}] with a gap between low and high LETs, $5.08 \leq LET_d \leq 10.8$. This gap appears because of lack of experimental data in intermediate domain of LET's. In low and high LETs two sets of linear and non-linear polynomials used. The fitted surfaces to the blue dots are the result of second / post-processing fitting procedure presented in this work that provides a continuous and smooth connection between low and high LET data sets within intermediate domain of LETs.
}
\label{Fig000}
\end{center}\vspace{-0.5cm}
\end{figure}

The numerical comparison of the proposed analytical model presented in this work and
the result of fitting to the clonogenic cell-survival data [\onlinecite{Guan2015:SR}] in low and high LETs are shown in Figure~\ref{Fig000}, and
series of Figures in Ref.[\onlinecite{Abolfath2017:SR}].

The fitting method is based on a three-parameter global fitting, e.g., a multi-variate modelling of cell survival~[\onlinecite{Akima1975:NIST}].
The numerical steps involve an optimization procedure that allows fitting of a 2D surface in a 3D parameter space spanned by dose, LET$_d$ and SF.
In this approach, the intrinsic correlations in cell survival data are incorporated into the optimization algorithm. As a result, background fluctuations of the extracted parameters $\alpha$ and $\beta$ of the linear quadratic model due to uncertainties in the biological data are considerably reduced.

The technical difference between the present fitting approach, shown in Figure~\ref{Fig000}, and similar figures shown in Ref.~[\onlinecite{Abolfath2017:SR}] is the way that we manipulated the extrapolation between low and high LET data points, in an intermediate domain of LET as described in the following.
In the present approach, based on the experimental points where LET $=$ 20 keV/$\mu$m is the maximum experimentally reported value, we divide the entire domain of LET into three regions of low, ${\rm LET}_d \leq 5.08 keV/\mu m$, high ${\rm LET}_d \geq 10.8 keV/\mu m$,  and intermediate LET.
The data points in low and high LET's are interpolated by a linear and nonlinear LET models as described in Ref.~[\onlinecite{Abolfath2017:SR}].

In a similar step, as described in Ref.~[\onlinecite{Abolfath2017:SR}], a numerical relation between cell survival, dose and LET has been extracted after  calculation of $\alpha$, $\beta$ and RBE where the fitting has been performed for each individual survival curve with a specific average LET.
However, because of lacking of the experimental data in the intermediate LET domain, i.e., $5.08 < {\rm LET}_d < 10.8$, and because the fitted surfaces in low and high LETs converge to different power law dependencies in ${\rm LET}_d$, any simplistic extrapolation of the data points from low and high to intermediate LET, causes a discontinuity in RBE and/or its derivatives.

To achieve a continuous and smooth RBE on the entire range of ${\rm LET}_d$, including in intermediate domain, in the present study we performed a post processing multivariate regression fitting model to connect the low- and high-LET RBE points.
The result of this fitting in 3D-space of SF, dose and LET is shown in Figure~\ref{Fig000}.
The calculated RBE shows a smooth and continuous transition from low to high LETs spite that there is a large gap in the experimental data within intermediate LET domain.

Notice that to obtain a smooth surface in extremely low doses we added $SF=1$ at $D=0$ for any LET to the experimental data-set and fitted the points. However, because of lack of experimental data in low doses, we cannot assess the accuracy of the fitted surface in extremely low dose limits.

The fitted RBE has been compared with several models in the literature including MKM~[\onlinecite{Hawkins1998:MP,Hawkins2003:RR,Kase2006}], LEM~[\onlinecite{22a,23a,24a,25a,27a}], linear models of Wilkens and Oelfke [\onlinecite{Wilkens2004:PMB}], Steinstr\"ater {\em et al.} [\onlinecite{Steinstrater2012:IJROBP}], and MCDS-RMF~[\onlinecite{Carlson2008:RR,Frese2012,Stewart2015}].
Figure 4 in Ref.[\onlinecite{Mohan2017:ARO}] shows a sample of such comparison between prediction of these methods and the experimental data.
Accordingly, these models are not able to explain the non-linearity of RBE at high LETs as reported by Guan {\em et al.}~[\onlinecite{Guan2015:SR}].
As shown in series of figures in Ref.~[\onlinecite{Abolfath2017:SR}], for low LET values, a linear dependence of $\alpha$ and constant $\beta$ on LET consistent with the fitting procedures introduced in other publications including in the above references have been found.
Contrary to these models, the present method is capable of capturing the non-linear increase of RBE as a function of LET, in particular in high LETs.
In particular, in domains beyond the Bragg peak where the stopping power reaches to its divergence point and the difference between the linearly fitted surface and the experimental data grows, the present model successfully captures smooth transition from low to high LET domains.
The details in computational steps and error analysis in fitting procedure of the present model including the goodness of the 3D global fit can be found in Refs.~[\onlinecite{Abolfath2016:MP,Abolfath2017:SR}].


Further inspection of the lineal-energy histograms as a function of track position in water reveals an abrupt transition in the energy loss spectrum in Bragg peak that resembles the aforementioned non-perturbative divergences in the series expansion representation of the biological response functions in high LET limits.
As demonstrated analytically in the preceding sections, the present study suggests that the difference between the data and the fitting models can be reduced by incorporating non-linear polynomials in $\alpha$, and $\beta$, calculated by taking into account three mechanisms due to (a) transition in the energy loss spectral density and considering a boundary that separates the spectrum of high energy particles from low energy particles, i.e., at Bragg peak, as well as the deviations of (b) energy loss from Poisson distribution in track structures and (c) chromosome aberrations from binary end-joining.

\begin{figure}
\begin{center}
\includegraphics[width=1.0\linewidth]{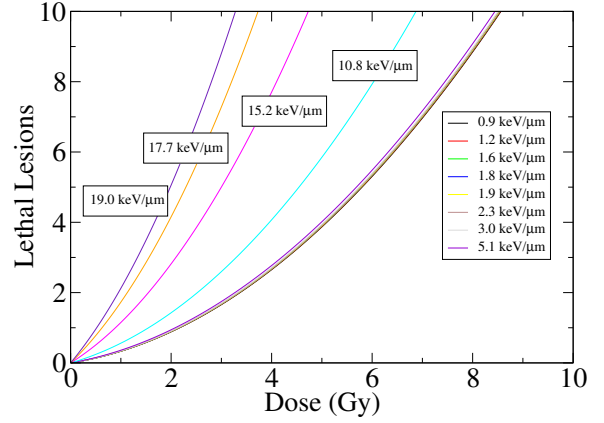}\\ 
\noindent
\caption{
Shown H460 
lethal lesions per cell as a function of dose and LETs.
The lethal lesions calculated using linear-quadratic model $\overline{L} = \alpha D + \beta D^2$.
The labels over each line represent experimental values of LET for specific lethal lesion curve.
}
\label{Fig0004}
\end{center}\vspace{-0.5cm}
\end{figure}

\begin{figure}
\begin{center}
\includegraphics[width=0.8\linewidth]{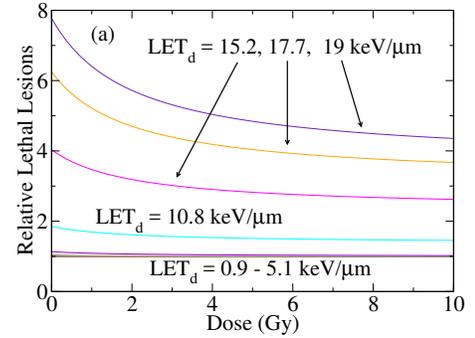}\\ 
\includegraphics[width=0.8\linewidth]{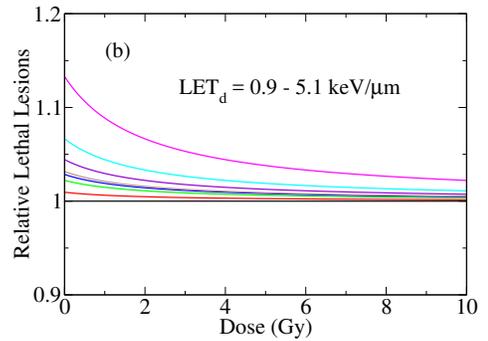}\\ 
\noindent
\caption{
Shown H460 (a and b) 
relative lethal lesions per cell as a function of dose for different LETs.
The relative lethal lesions, $\overline{L}(LET_d)/\overline{L}(LET_d=0.9)$, were calculated using 3D global fitting of the experimental cell survival data within LQ model.
It represents $\overline{L}(LET_d)$, normalized by the lethal lesions at the lowest LET, $\overline{L}(LET_d=0.9)$.
For a better visualization of relative lethal lesions in low LETs, a magnified version of (a) is shown in (b).
}
\label{Fig0005}
\end{center}\vspace{-0.5cm}
\end{figure}

\subsection{Lethal lesions}
We now turn to investigate the prediction of the present model, i.e., Eq.~(\ref{A13}), on the lethal lesions as a function of absorbed dose and LET.
Figures \ref{Fig0004} and \ref{Fig0005} show the lethal and relative lethal lesions, respectively, as a function of absorbed dose for different LETs, calculated by LQ model $\overline{L} = \alpha D + \beta D^2$.
As introduced in Refs. [\onlinecite{Abolfath2016:MP,Abolfath2017:SR}] and described in this work, the experimental values for $\alpha$ and $\beta$ were calculated from a polynomial expansion and 3D global fitting approach to obtain the best-fitted surface to the experimental data [\onlinecite{Guan2015:SR}].
As seen in Figure \ref{Fig0004}, the present model predicts a substantial increase in lethal lesions in LET's corresponding to the domains close to and beyond the Bragg peak where $LET_d \geq 10.8 keV/\mu m$.

Similarly, the results of our calculation on the relative lethal lesions presented in Figure \ref{Fig0005} where a normalization to the lethal lesions corresponding to the lowest experimentally accessible LET, e.g., $LET_d = 0.9 keV/\mu m$ has been performed.
Interestingly, as seen in Figure \ref{Fig0005}(a), in this scale, the relative lethal lesions as a function of dose in low LET's resemble straight lines coincide on the top each other.
To distinguish the differences between curves corresponding to low LET's, we magnified and depicted low LET curves in Figure \ref{Fig0005}(b).

Hence our model predicts small variation in relative lethal lesions in a wide interval of beam range, e.g., from the beam entrance to the Bragg peak.
This is in contrast to the narrow range of distal points to the Bragg peak, e.g., high LET's, that the lethal lesions increase abruptly, a manifestation of the non-linearity in RBE.
As expected, in low doses, where $\overline{L}$ scales linearly with deposited dose, the relative lethal lesions starts from $\overline{L}(LET_d) / \overline{L}(LET_d=0.9) = \alpha(LET_d)/\alpha(LET_d=0.9)$ where $\overline{L}(LET_d) / \overline{L}(LET_d=0.9)$ is the maximum, as $\overline{L}$ monotonically increases with the increase of $LET_d$~[\onlinecite{Abolfath2016:MP,Abolfath2017:SR}].
With increase in the deposition dose, $\overline{L}(LET_d) / \overline{L}(LET_d=0.9)$ drops and approaches to a saturation value, given by
$\overline{L}(LET_d) / \overline{L}(LET_d=0.9) = \beta(LET_d)/\beta(LET_d=0.9)$.
Similar to $\alpha$, the 3D global fitting of the experimental data has shown that $\beta$ increases with the increase of $LET_d$, but with slower rate such that $\alpha/\beta > 1$ hence $\overline{L}(LET_d) / \overline{L}(LET_d=0.9)\mid_{D\rightarrow 0} > \overline{L}(LET_d) / \overline{L}(LET_d=0.9)\mid_{D\rightarrow \infty}$.

A simplistic interpretation of the lethality as a function of dose and LET, as shown in Figure~\ref{Fig0005}, indicates that as dose increases, the dependence of biological effect (relative number of lesions, complexity of damage, etc.) on LET decreases. More specifically, because of dominance of $\beta$ in cell-survival in high doses and the fact that $\alpha/\beta$ is an increasing function of LET (see Ref.~[\onlinecite{Abolfath2017:SR}] for details on dependence of $\alpha/\beta$ on LET), the stronger LET dependence, originated from $\alpha$ diminishes in favor of weaker LET dependence in $\beta$. Therefore a decline in lethality as an increasing function of dose for a given LET is expected.

Note that in models such as MKM, and MCDS-RMF where $\beta$ is assumed to be independent of $LET_d$, the relative lethal lesions in high doses is predicted to approach to unity independent of $LET_d$, e.g., $\overline{L}(LET_d) / \overline{L}(LET_d=0.9)\mid_{D\rightarrow \infty} = 1$.
According to the predictions of the present model, by counting the number of lethal lesions, $\gamma$-H2AX or Foci, and their asymptotic behavior in high doses, one can determine the dependence of $\beta$ on $LET_d$ from an independent measurement~[\onlinecite{BronkAAPM:2017}].

Recently, in an unpublished dataset, Bronk et al. reported a dose and LET-dependent measurement of DNA DSB repair Foci above endogenous background which persist for 24 hours following irradiation~[\onlinecite{BronkAAPM:2017}]. Similar to the predicted lethal lesions, as shown in Figures \ref{Fig0004} and \ref{Fig0005}, the ``persistent'' Foci show a rapid increase near the Bragg peak and beyond. The experimentally selected persistent Foci are a subclass of Foci which presumably correspond to single hit hence their counts increases linearly in dose with a slope proportional to $\alpha$.
The prediction of the current model on such lethal lesions, modeled by $\overline{L} \propto \alpha D$ is qualitatively in agreement with the experiment. 
One of the important outcomes of this measurement is the divergence of lethal lesion lines corresponding to different $LET_d$'s in high doses that indicates the dependence of $\beta$ on $LET_d$.
Further studies are on the way to converge the theoretical predictions with experiments.

\subsection{Tumor control probability}
To incorporate the LET dependence of TCP, we first start with a single voxel.
In general such TCP depends on cell-line, number of embedded tumor cells in voxel, dose, and LET and number of treatment fractions.
Considering a voxel with label $k$ in a tumor with an average number of tumor cells, $N_k$,
we formally calculate a single fraction TCP by
\begin{eqnarray}
TCP_k = (1 - SF_k)^{N_k},
\end{eqnarray}
where ${\rm SF}_k = \exp(-\alpha(LET_k) D_k -\beta(LET_k) D_k^2)$ is the $10\%$ cell survival fraction. 
The dependence of TCP$_k$ on LET and dose comes from the average value of LET and dose in the voxel, LET$_k$ and $D_k$.
Hence TCP over the entire volume of the tumor covered by a single scanning proton beam can be calculated by an integration over the voxels under radiation.
Considering Poisson model for TCP where $SF << 1$ [\onlinecite{ZaiderPMB:2000}], it follows
\begin{eqnarray}
{\rm TCP} &=& \prod_k {\rm TCP}_k = \prod_k (1 - {\rm SF}_k)^{N_k} \\ \nonumber
&\approx& \prod_k e^{- N_k \times {\rm SF}_k} = e^{- \sum_k N_k \times {\rm SF}_k}.
\end{eqnarray}
The above summation can simply change to integration over the corresponding target volume, $V$
\begin{eqnarray}
- \frac{\ln({\rm TCP})}{N_v} = \int_{V} \frac{d\vec{r}}{\Omega_v} e^{-\alpha(LET_{\vec{r}}) D -\beta(LET_{\vec{r}}) D^2},
\label{tcp}
\end{eqnarray}
where $N_v$ is the average number of cells in a single voxel with volume $\Omega_v$. The TCP calculated by Eq.~(\ref{tcp}), depends only on dose as the dependence over lineal-energy and LET has been integrated out.

\begin{figure}
\begin{center}
\includegraphics[width=1.0\linewidth]{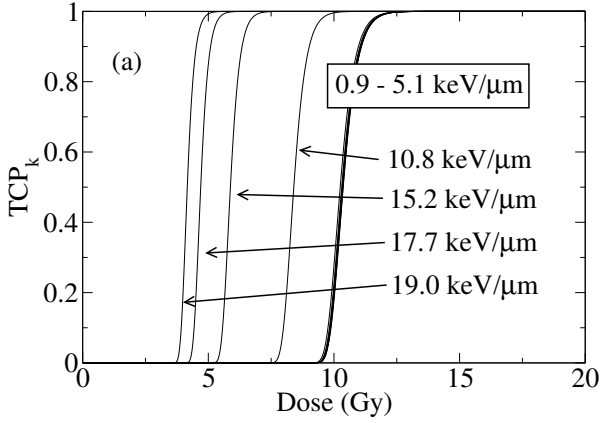}\\ \vspace{-0.25cm}
\noindent
\caption{
Shown the tumor control probability, TCP, of $10^6$ H460 NSCLC in a typical mm-size volume.
As the number of cells in the volume increases the sigmoid curve shifts to higher doses.
}
\label{Fig0007}
\end{center}\vspace{-0.5cm}
\end{figure}

We now turn to present an estimate on TCP for a tumor consists of H460 NSCLC cells where the dependence of SF on dose and LET measured experimentally in our group and fitted by 3D global fitting method by the present authors.
Depends on the location of voxel in beam, $TCP_k$ changes from high doses to low doses as LET increases.
As shown in Figure \ref{Fig0007}, TCP does not changes significantly with LET up to an LET value of $LET_d = 5.1$ keV/$\mu m$, which corresponds to a depth proximal to the Bragg peak.
Within this range of LETs, all TCP curves coincide on the top each other.
Close and beyond the Bragg peak we observe a drastic change in TCP, where increasing LET lowers radiation dose with a specific TCP (e.g., $TCP_D = 50\%$) abruptly.


\begin{figure}
\begin{center}
\includegraphics[width=1.0\linewidth]{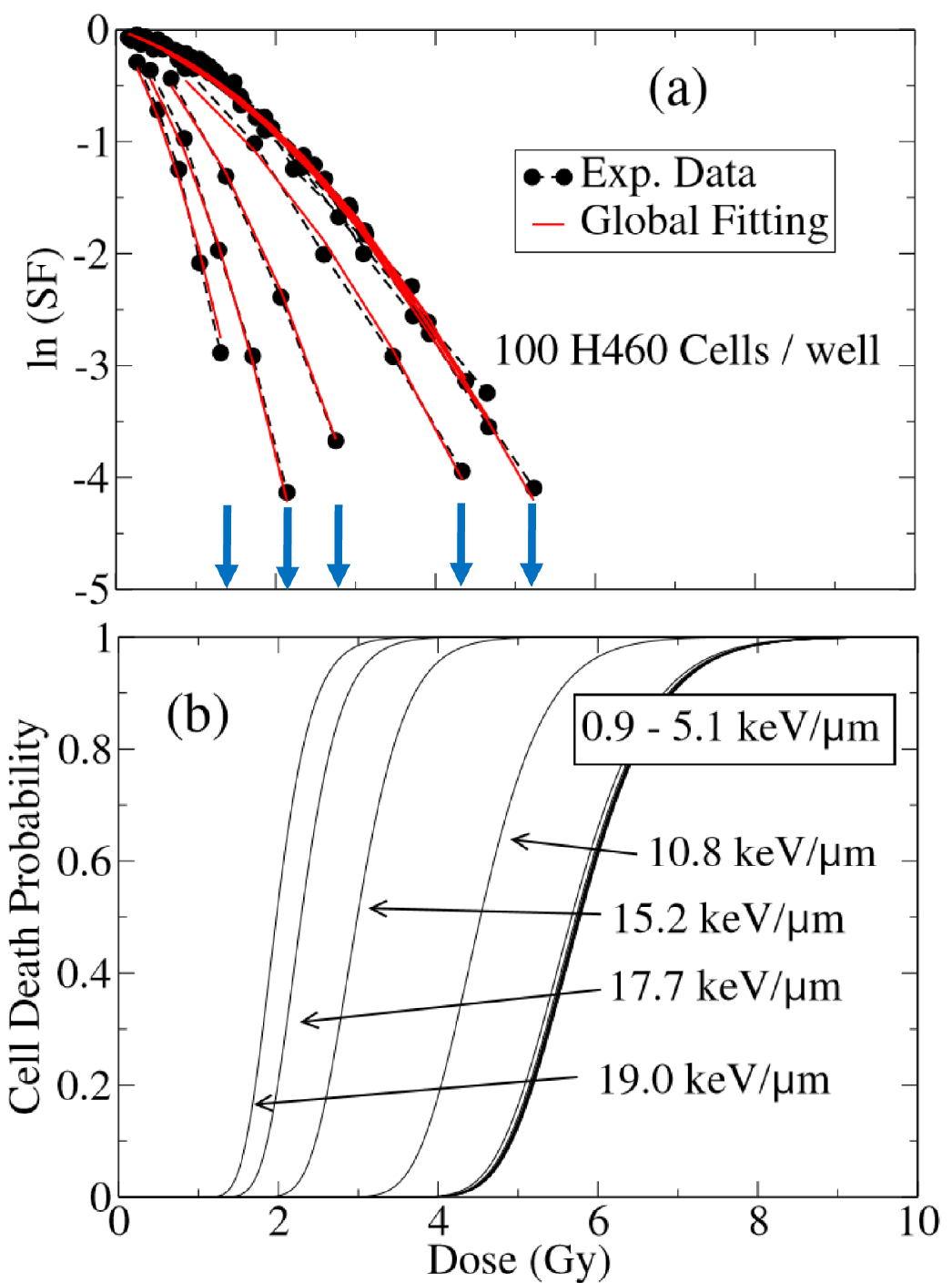}\\ \vspace{-0.25cm}
\noindent
\caption{
Shown (a) experimental (circles) fitted by three-dimensional global fitting (red bold lines) of cell survival fraction (SF) and (b) the cell death probability (CDP) of 100 H460 NSCLC used experimentally to measure SF in a 12 wells-plate as a function of dose and LET, irradiated by a single fraction of scanning proton with approximate nominal energy of 80 MeV.
The end points in SF where the radiation dose is highest, are indicated by arrows in (a), coincide with the onset of CDP sigmoid turning points.
A sharp rise in sigmoid justifies termination in viability of cells seeded in the well, hence, a drop in their biological responses.
With increase in LET, the maximum experimentally achievable doses in cell viability that terminate SF curves, lowers.
}
\label{Fig8}
\end{center}\vspace{-0.5cm}
\end{figure}

\subsection{Cell death probability and in-vitro SF}
Similar to TCP, we define cell death probability (CDP) for much smaller number of cells collected and sampled for in-vitro assays.
In our latest high-throughput experiments, 100 cells per well have been seeded and for each dose-LET combination 16 experimental samples were replicated.
In this series of experiments, the number of cells per colony after irradiation depends on the dose and LET. Within each condition there is a distribution of cells and colonies.
We therefore use same concepts and formulas of TCP to calculate CDP for entire cell population death in wells/plates as function of dose and LET.

The results of SF data collected from these experiments on H460 NSCLC cells is summarized and shown in Figure~\ref{Fig8}.
In general as we discussed in formulation of the complexity in cell damage in the preceding sections, the more complex/thorough damage should in theory result in a longer delay/cell cycle arrest and thus less cells in colonies which managed to eventually repair the damage to go on to form a viable colony. Essentially, the colony morphology distribution appears to become more heterogeneous/broad with increasing damage.

A hallmark of these experiments is systematic elevation in cell-death with increases in LET.
In particular, the maximum measured doses that lead to entire cell death in a well in higher LET's, corresponding to last circles in Figure~\ref{Fig8}(a) where SF curves are terminated, are appeared in lower values.
For example the entire cell population in a well exposed by 80 MeV proton beam with $LET=19 keV/\mu m$, did not survive beyond a dose of $D=1.6$ Gy.
In Figure~\ref{Fig8}(a), this point corresponds to the last circular point in the first curve from left hand side.

This observation has raised questions on computation of the detection limits and possibly saturations in cell death around empirically observed maximum doses and if such calculation would be useful in providing rough estimates on TCP in larger and denser population of cells such as in tumors, assuming similar in-vivo biological conditions as in in-vitro biological endpoints.

Interestingly enough, the three-dimensional global fitting integrated into SF and CDP calculation shows occurrence of a saturation around specific doses and LET's.
These points match with a series of maximum doses given by specific LET's reported experimentally, and clearly visible in Figure~\ref{Fig8}.
Similar to TCP calculation, we consider irradiation of cells by 80 MeV scanning beam of protons delivered in a single fraction, and
use a linear-quadratic model to calculate CDP.
The numerical values for $\alpha$ and $\beta$ were calculated based on a polynomial expansion in three-dimensional global fitting of SF, dose and LET from the reported experimental data in our group.

In Figure~\ref{Fig8}(a), the experimental points are shown in circles and the red solid lines are the result of 3D global fitting.
For each individual LET, there is a maximum dose at which the data set is terminated, indicating that the cell survival detection limit has been reached because of inactivity of majority of the seeded cells.
The blue arrows are depicted in Figure~\ref{Fig8}(a) to show the series of maximally achievable doses for each LET.
In Figure~\ref{Fig8}(b) we show the result of CDP calculation as a function of dose and LET for a sample of 100 H460 NSCLC, using same scale of dose as used for $x$-axis in Figure~\ref{Fig8}(a).
For illustration of the physical picture discussed above, let us consider our previous example of SF curve corresponding to $LET=19 keV/\mu m$ in Figure~\ref{Fig8}(a) and draw a vertical line at $D=1.6$ Gy, starting from the last circle in the curve, down to the horizontal axis in Figure~\ref{Fig8}(b). This line connects $D=1.6$ Gy in  Figure~\ref{Fig8}(a) and (b) and hit a point in the shoulder of sigmoid in CDP.
By repeating this for the entire LET SF curves, we observe a patter as discussed here.
Hence, the end points in SF where the radiation dose is highest, coincide with the onset of CDP sigmoid turning points.
Sharp rises in sigmoid resemble a rapid termination in the measured biological activities of the seeded cells.
Similar to TCP, depicted in Figure~\ref{Fig0007}, with an increase in LET, the threshold in cell death falls to lower doses.
Comparison with the experimental data confirms a shift in sigmoid turning points to higher doses as the number of seeded cells increases.

\section{Discussion and conclusion}
The cell survival data reported in Guan {\it et al.}[\onlinecite{Guan2015:SR}] show the measured biologic effects are substantially greater than in most previous reports. It is characterized by a non-linear RBE as a function of LET near and beyond the Bragg peak. The calculated RBE is characterized by high sensitivity to small variations in LET distal to the Bragg peak, where a small uncertainty in the position of the cells may result in significant change in LET.

Extensive efforts in achieving agreement between the current formulations and approximations of the standard RBE models including MKM, LEM, and MCDS-RMF resulted in no success where all of these models appeared to underestimate RBE of the experimental data~[\onlinecite{Guan2015:SR}] in high LET's~[\onlinecite{Mohan2017:ARO}].
It is therefore crucial to search for appropriate fitting procedures and models to be able to enhance the quality of the calculated RBE and the interpretation of the data.

To address these issues, we recently engaged different approach and fitted SF-dose-LET experimental data using polynomial expansion of $\alpha$ and $\beta$ as a function of LET$_d$~[\onlinecite{Abolfath2017:SR}].
We continued with this computational model, and in this work (1) introduced a detailed analytical analysis and developed a mathematical framework and built upon event-by-event track structure simulations and microdosimetry and its formalism to analyze and interpret the biophysical processes relevant to each term in the polynomial expansion and (2) improved the quality of the fitting procedure.

In addition to initial DNA damage, chromatin dynamics were incorporated in our model through phenomenological repair and mis-repair rate equations.
We focused on two major classes of deviations from the standard RBE models in high LET domains, e.g., in distal edge to the Bragg peak. These mechanisms are mainly deviation of energy loss from Poisson distribution in track structures discussed in Section \ref{sect_1} and chromosome aberrations from binary end-joining discussed in Section \ref{sect_2}.
We specifically predicted occurrence of continuous transitions in population of chromosome aberration complexities amongst binary, ternary, quaternary and higher order combinations as a function of proton LET, a hypothesis that must be verified experimentally.

We illustrated the contribution from the energy loss fluctuations renormalizes the LQ biological parameters $\alpha $ and $\beta $ to infinite orders in $z_D$ and $y_D$.
Interplay between $y_D$ and LET$_d$, obtained in this work from Monte Carlo simulations, makes a bridged between the microscopic models and macroscopic observations and experimental data.
In particular we demonstrated a non-linear dependence between $y_D$ and LET$_d$, that emerges around ${\rm LET}_d \approx 5 keV/\mu m$,
attributes to a universal non-linearity in radio-biological index $\alpha$ even if $\alpha$ scales linearly with $y_D$ as predicted in typical radiobiological models.

We furthermore improved our methodology and results on global fitting algorithm of the biological responses of cells under exposure of scanning beam of protons, recently presented in Ref.~[\onlinecite{Abolfath2017:SR}] where a linear function of $\alpha$ on LET$_d$ and constant $\beta$ fitted the experimental data of two types of NSCLC H460 and H1437 cell lines reported by Guan et al. ~[\onlinecite{Guan2015:SR}] in low LETs.
In high LET's, the deviation from linear LET model appeared to be prominent such that a non-perturbative correction to the linear model would be essential.
Thus analytical model presented in this work reveals several complex aspects and non-trivial predictions of such non-linearity, emerging as LET increases monotonically.

Specific to these experimental data~[\onlinecite{Guan2015:SR}], we divided LET into three domains of low, intermediate and high, ${\rm LET}_d \leq 5.08 keV/\mu m$, $5.08 < {\rm LET}_d < 10.8 keV/\mu m$, and ${\rm LET}_d \geq 10.8 keV/\mu m$ respectively.
In our previous work~[\onlinecite{Abolfath2017:SR}], we performed fitting of the experimental data to linear and non-linear functions of LET in low and high LET domains separately.
In this work, we connected these solutions numerically by
incorporating matching of the low and high LET solutions in intermediate domain of LETs where no experimental data were reported.
This extra step of matching allows obtaining continuous SF-dose-LET surfaces and RBE and their higher gradients for the entire domains of LET.
Otherwise, a discontinuity in RBE and/or its higher derivatives would appear as a numerical artifact in the intermediate region because of differences in order of polynomials used for fittings in low and high LETs.
To this end, in the present study, we performed a post processing multivariate regression fitting algorithm~[\onlinecite{Akima1975:NIST}] as well as a
one-dimensional smooth interpolation~[\onlinecite{Akima1970:JACM}] and obtained a continuous RBE, $\alpha$, $\beta$ and $\alpha/\beta$ curves and their higher derivatives.
Within a range of LETs reported in the experiment in Ref.~[\onlinecite{Guan2015:SR}], these parameters grow continuously as LET increases.
A monotonic trend in lethal lesions, $\alpha$, $\beta$ and $\alpha/\beta$ can be anticipated via a subclass of the target theories, where $\alpha$ and $\beta$ represent DNA damage induction by one- and two-track action of ionizing radiation that leads to cell inactivation~[\onlinecite{Vassiliev2012:IJROBP,Vassiliev2017:PMB}].


We then applied the fitted results to the standard TCP and CDP models and reported theoretical predictions of these parameters based on the fitting of the experimental data in Ref.~[\onlinecite{Guan2015:SR}].
We compared the rise in CDP, the onset of the shoulders between plateaus in sigmoid model of CDP, with in-vitro experimental data and identified the characteristics of the maximally achievable absorbed doses as a function of LET beyond which no biological activity in the wells and plates were observed.
We found a good agreement between model calculation as well as physical interpretation of the empirical observations.

We now turn to remark on the implications of the presented TCP results in clinical applications of proton therapy as shown in Figure \ref{Fig0005}.
In particular in low doses where the induction of the lethal lesions is significantly higher in high LET's,
consistent with monotonic increase of $\alpha/\beta$ as a function of LET.
To be more specific, let us consider a normal tissue or an organ at risk (OAR), located posterior to a tumor, where the beam exits.
In an optimized dose distribution plan, where the scanning spot patterns cover the entire volume of the target, the Bragg peak distal edge falls in posterior edge of the target beyond which the dose drops rapidly.
Let us also consider a hypothetical situation in which proton deposited dose in a normal tissue to be within a factor of tenth relative to the average dose in the tumor.
In the present clinical setup where the dose distribution and the dose volume histograms are the geometrical metrics in evaluating a patient plan, a tenth of prescribed dose in a normal organ can be considered within the tissue tolerance.
However, if we take into account the cell toxicities by LET, the dose weighted by relative lethal lesions as illustrated in Figure \ref{Fig0005},
substantially changes the predictive outcome since the relative number of lethal lesions per Gy per cell in such volumes can be as high as a factor of ten.
Moreover because of monotonic increase of LET and lethal lesions as shown in Figure \ref{Fig0005}, the cell toxicity posterior to the tumor would be significantly different from its anterior side.

In treatment modalities by photon, it is conventional to avoid OARs and normal tissues at the beam entrance, anterior to the tumor.
Instead, treatment planners may allow OARs to be in locations at the exit dose to lower the toxicity.
In treatments using scanning beam of proton, however, if we take the biological responses such as the ones depicted in Figure \ref{Fig0005} into account,
we may anticipate a drop in physical dose may cancel out an increase in cell toxicity, in particular in tissue volumes where the exit dose drops to approximately to 1/10 of the dose compare to beam entrance.
Beyond these points, the drop in dose can no longer be compensated by increase in lethal lesions, hence the net effect of dose times RBE falls rapidly, unless if the number of secondary particles would be significantly large because of a particular beam arrangement.

The above arguments rationalize a need in extending to generally expected rules in plan evaluation techniques that are currently based on only the physical dose, e.g., the beam conformity index, dose volume histogram (DVH) and etc., to include accurately the biological effects and proton LET's in TCP and normal tissue complication probability (NTCP).
Clearly a more reliable clinical plan evaluation must rely on convolution of the physical dose and cell toxicity and biological responses.

Similarly for tumors, it is customary to apply a uniform dose distribution in conventional radiation therapy by photons.
However, incorporating LET to the dose in TCP shows a need in tailoring this approach.
In particular volumes irradiated by high LET protons, the optimal dose from TCP is expected to be lower compare to volumes irradiated by low LET protons.
Hence to achieve a uniform TCP a non-uniform dose distribution should be delivered to the target.
To this end, we presented a calculation of TCP to account for spatial variation of LET and prescribed dose in treatments by scanning beam of protons.
Our calculation shows that $TCP$ will be inhomogeneous if a homogeneous prescribed dose is applied. Conversely, one may optimize an inhomogeneous dose distribution in the target to achieve a homogeneous $TCP$.

Unlike the prescription dose in treatments by photons where a homogenous radiation dose is desired, in scanning proton therapy, and in general, particle therapy, the effect of LET prevents application of a uniform dose throughout the target volume.
This phenomenon knows as LET-painting has been recently investigated for treatment of tumors with hypoxic conditions~[\onlinecite{Tinganelli2015:SR,Kramer2014:EPJD,Bassler2014:AO}].
For example for small targets, such as metastatic brain tumors, where scanning beam of proton with a single energy can be used, a given TCP across the target volume requires prescribing smaller dose toward the exit points to compensate the effect of higher LET's.
In this case, a distribution of beam angles, e.g., a conformal arc, may help spreading LETs across the edge of the target to achieve a uniform TCP with application of a uniform deposited dose.

We finally remark that the complex aspects discussed above can not be anticipated and captured only by performing an adhoc Taylor and/or power expansion of $\alpha$ and $\beta$ as a function of LET, a practical approach only useful for mathematical fitting of the experimental data.
This clearly justifies a need for developing step by step derivation of equations and sophisticated algebra as performed in this work.
Thus our approach in combining numerical fitting procedure with analytical track structure provides a systematic path to apply relevant corrections to the current RBE models to be able to reconcile theory with the recent experimental data.

An extension of this approach for particles heavier than protons is currently under investigation and the goal for such studies is to generate data needed to optimize treatment plans incorporating the variable RBE.

\section{Appendix A: derivation of DSB rate equation}
\label{AD}
In this Appendix we present the derivation of Eq.~(\ref{eq0025A}) and show $d\overline{n}(t)/dt = \sum_{n=0}^\infty (g_n - r_n) Q_n(t),$ is the rate equation associated with the master equation, $dQ_n(t)/dt = g_{n-1} Q_{n-1} - g_n Q_n + r_{n+1} Q_{n+1} - r_n Q_n,$ where $\overline{n}(t)=\sum_{n=0}^\infty n Q_n(t)$, hence
\begin{eqnarray}
\frac{d\overline{n}(t)}{dt} &=& \sum_{n=0}^\infty n \frac{dQ_n(t)}{dt} \nonumber \\
&=& \sum_{n=0}^\infty n (g_{n-1} Q_{n-1} - g_n Q_n  \nonumber \\ &&
+ r_{n+1} Q_{n+1} - r_n Q_n).
\label{eq0020}
\end{eqnarray}
At the boundary of Markov chain, $n=0$, the series run over $n=-1$, as seen in Eq.~(\ref{eq0020}), and sketched in Figure~\ref{Fig00}.
To resolve this issue, we extend the chain to include negative $n$'s, but with probability $Q_{-|n|}=0$.
Also note that the DNA initial condition asserts an assumption on post-irradiation repair mechanisms. Because the DNA-DSB dynamics starts from acute irradiation that takes place at $n=0$, e.g., the state of DNA with zero DSB, we assume $r_{n=0}=0$.
To this end, we simplify Eq.~(\ref{eq0020}), by splitting the series into
\begin{eqnarray}
\frac{d\overline{n}(t)}{dt} = \frac{d\overline{n}_g(t)}{dt} + \frac{d\overline{n}_r(t)}{dt},
\label{eq0021}
\end{eqnarray}
where
\begin{eqnarray}
&&\frac{d\overline{n}_g(t)}{dt} = \sum_{n=0}^\infty n \left(g_{n-1} Q_{n-1} - g_n Q_n\right) \nonumber \\ &&
= 0 \times (g_{-1} Q_{-1} - g_0 Q_0) + 1 \times (g_0 Q_0 - g_1 Q_1) \nonumber \\ &&
+ 2 \times (g_1 Q_1 - g_2 Q_2) + 3 \times (g_2 Q_2 - g_3 Q_3) \nonumber \\ &&
+ 4 \times (g_3 Q_3 - g_4 Q_4) + \dots \nonumber \\ &&
= g_0 Q_0 + g_1 (2 Q_1 - Q_1) + g_2 (3 Q_2 - 2 Q_2) \nonumber \\ &&
+ g_3 (4 Q_3 - 3 Q_3) + \dots \nonumber \\ &&
= g_0 Q_0 + g_1 Q_1 + g_2 Q_2 + \dots
\label{eq0022}
\end{eqnarray}
and thus
\begin{eqnarray}
&&\frac{d\overline{n}_g(t)}{dt} = \sum_{n=0}^\infty g_n Q_n.
\label{eq0023}
\end{eqnarray}
Similarly
\begin{eqnarray}
&&\frac{d\overline{n}_r(t)}{dt} = \sum_{n=0}^\infty n \left(r_{n+1} Q_{n+1} - r_n Q_n\right) \nonumber \\ &&
= 0 \times (r_1 Q_1 - r_0 Q_0) + 1 \times (r_2 Q_2 - r_1 Q_1) \nonumber \\ &&
+ 2 \times (r_3 Q_3 - r_2 Q_2) + 3 \times (r_4 Q_4 - r_3 Q_3) \nonumber \\ &&
+ 4 \times (r_5 Q_5 - r_4 Q_4) + \dots \nonumber \\ &&
= - r_1 Q_1 - r_2 (2 Q_2 - Q_2) - r_3 (3 Q_3 - 2 Q_3) \nonumber \\ &&
- r_4 (4 Q_4 - 3 Q_4) + \dots \nonumber \\ &&
= - r_1 Q_1 - r_2 Q_2 - r_3 Q_3 + \dots
\label{eq0024}
\end{eqnarray}
and thus
\begin{eqnarray}
&&\frac{d\overline{n}_r(t)}{dt} = - \sum_{n=1}^\infty r_n Q_n = - \sum_{n=0}^\infty r_n Q_n.
\label{eq0025}
\end{eqnarray}
In the last equation we added $r_0 Q_0$ to the series as $r_0 = 0$.
Combining Eqs.~\ref{eq0021},~\ref{eq0023}, ~\ref{eq0025} we end up with derivation of Eq.~(\ref{eq0025A})
\begin{eqnarray}
\frac{d\overline{n}(t)}{dt} = \sum_{n=0}^\infty (g_n - r_n) Q_n(t).
\label{eq0026}
\end{eqnarray}

\section{Appendix B: Lineal energy or LET representations}
In this Appendix, we present our numerical approach in simulating moments of specific and lineal energies, LET's and their spectrum
by performing MC sampling of event by event energy transfers and stepping lengths in  track structures.
We begin with numerical construction of the lineal energy spectral density that used in the present study and our recent publication, Ref.~[\onlinecite{Abolfath2017:SR}].
Samples of lineal energies scored by MC in a specific target volume and sorted in ascending order.
A uniform bin size through the entire range of lineal-energies were defined and the number of lineal energies in successive intervals, $\Delta y$, were enumerated.
Subsequently statistical moments of specific and lineal energies were calculated.

To this end, in a target volume and/or voxel, labeled  by index, $j$, we scored the energy transfers and the associated stepping track length at point of interaction, $i$, for a particle with species, $\sigma$, and constructed a random matrix from lineal energy elements $y_{\sigma i, j} = \left(d\varepsilon_{\sigma i}/dl_{\sigma i}\right)_j$.
We subsequently calculated the distribution function using the following formulation
\begin{eqnarray}
S_j(y) = \frac{1}{N_{y,j}} \sum_{\sigma i} \delta(y - y_{\sigma i, j}).
\end{eqnarray}
$S$ is normalized in each individual voxel such that
\begin{eqnarray}
1_j &=& \int_{-\infty}^{\infty} dy S_j(y) \nonumber \\
&=& \int_{-\infty}^{\infty} dy \frac{1}{N_{y,j}} \sum_{\sigma i} \delta(y - y_{\sigma i, j})  \nonumber \\
&=& \frac{1}{N_{y,j}} \sum_{\sigma i} \int_{-\infty}^{\infty} dy \delta(y - y_{\sigma i, j})  \nonumber \\
&=& \frac{1}{N_{y,j}} \sum_{\sigma i} 1_{\sigma i, j}.
\end{eqnarray}
Here $\sum_{\sigma i} 1_{\sigma i, j} = N_{y,j}$ represents number of lineal energy events.
The average value of lineal energy is then given by
\begin{eqnarray}
\overline{y}_j &=& \int_{-\infty}^{\infty} dy y S_j(y) \nonumber \\
&=& \int_{-\infty}^{\infty} dy y \frac{1}{N_{y,j}} \sum_{\sigma,i} \delta(y - y_{\sigma i, j}) \nonumber \\
&=& \frac{1}{N_{y,j}} \sum_{\sigma,i} \int_{-\infty}^{\infty} dy y \delta(y - y_{\sigma i, j}) \nonumber \\
&=& \frac{1}{N_{y,j}} \sum_{\sigma i} y_{\sigma i, j}.
\label{eq44}
\end{eqnarray}
Here $\sum_{\sigma i} y_{\sigma i, j}$ is algebraic sum of lineal energies in a voxel.
By repeating the above procedure we can calculate the higher order statistical moments of lineal energy, as given below
\begin{eqnarray}
\overline{y^n}_{j} = \int_{-\infty}^{\infty} dy y^n S_j(y) = \frac{1}{N_{y,j}} \sum_{\sigma i} y^n_{\sigma i, j},
\end{eqnarray}
where $n=1,2,3, ...$.

Alternatively we may calculate a $y$-averaged LET (or $y_D$)
\begin{eqnarray}
y_{D,j} = {\rm LET}_{y,j} 
&=& \left(\sum_{\sigma i} \frac{d\varepsilon_{\sigma i}}{dl_{\sigma i}} Y_{\sigma i}\right)_j \nonumber \\
&=& \left(\frac{\sum_{\sigma i} d\varepsilon_{\sigma i}/dl_{\sigma i} \times (d\varepsilon_{\sigma i}/dl_{\sigma i})}{\sum_{\sigma i} d\varepsilon_{\sigma i}/dl_{\sigma i}}\right)_j \nonumber \\
&=& \left(\frac{\sum_{\sigma} \int_0^\infty dE \Phi_\sigma(E)~ y^2_{\sigma}(E)}{\sum_{\sigma} \int_0^\infty dE \Phi_\sigma(E)~ y_{\sigma}(E)}\right)_j,
\label{eq1_34}
\end{eqnarray}
where $\sum_{\sigma i} \rightarrow \sum_{\sigma}\int_0^\infty dE \Phi_\sigma(E)$.
Here $\Phi_\sigma(E)$ is the differential particle fluence of species $\sigma$ with kinetic energy $E$ and $Y_{\sigma i}$ denotes the spectral density matrix with elements containing randomly generated lineal-energies
\begin{eqnarray}
Y_{\sigma i} = \frac{d\varepsilon_{\sigma i}/dl_{\sigma i}}{\sum_{\sigma i} d\varepsilon_{\sigma i}/dl_{\sigma i}}.
\label{eq1_35}
\end{eqnarray}
Note that $Y$ is a dimensionless distribution function where $S$ is a histogram that counts number of lineal-energy per interval $\Delta y$ as defined above.
We call ${\rm LET}_{y,j}$, $y$-averaged LET.

To be consistent with the distribution function used for calculation of LET, one may define a deposited dose, averaging over the same distribution function.
For example of $y$-representation of distribution function
\begin{eqnarray}
{\rm D}_{y,j} &=& \sum_{\sigma i} z_{\sigma i} Y_{\sigma i} \nonumber \\
&=& \left(\frac{\sum_{\sigma i} (d\varepsilon_{\sigma i}/dm_j) \times (d\varepsilon_{\sigma i}/dl_{\sigma i})}{\sum_{\sigma i} d\varepsilon_{\sigma i}/dl_{\sigma i}}\right)_j,
\label{eq1_5}
\end{eqnarray}
where $dm$ is element of the mass in voxel $j$th.

For completeness of our presentation, we briefly discuss the construction of dose and track averaged LET, although they were heavily presented in literature (see, e.g., ~[\onlinecite{Guan2015:MP}] and the references therein).
In calculation of dose-averaged LET, ${\rm LET}_{d,j}$, we use the following spectral density
\begin{eqnarray}
Z_{\sigma i} = \frac{d\varepsilon_{\sigma i}}{\sum_{\sigma i} d\varepsilon_{\sigma i}},
\label{eq1_35}
\end{eqnarray}
hence
\begin{eqnarray}
{\rm LET}_{d,j} 
&=& \left(\sum_{\sigma i} \frac{d\varepsilon_{\sigma i}}{dl_{\sigma i}} Z_{\sigma i}\right)_j \nonumber \\
&=& \left(\frac{\sum_{\sigma i} d\varepsilon_{\sigma i} \times (d\varepsilon_{\sigma i}/dl_{\sigma i})}{\sum_{\sigma i} d\varepsilon_{\sigma i}}\right)_j
\nonumber \\
&=& \left(\frac{\sum_{\sigma} \int_0^\infty dE \Phi_\sigma(E)~ z_{\sigma}(E) \times y_{\sigma}(E)}{\sum_{\sigma} \int_0^\infty dE \Phi_\sigma(E)~ z_{\sigma}(E)}\right)_j,
\label{eq1_3}
\end{eqnarray}
where $d\varepsilon_{\sigma i} \rightarrow d\varepsilon_{\sigma i}/dm = z_{\sigma}(E)$. 
Note that $d\varepsilon_i/\sum_i d\varepsilon_i$ is a normalized distribution function of the energy transfer in a given voxel, hence any dose averaged quantity, including the deposited dose, in this representation, were calculated by the following equation
\begin{eqnarray}
{\rm D}_{d,j} &=& \sum_{\sigma i} z_{\sigma i} Z_{\sigma i} \nonumber \\
&=& \left(\frac{\sum_{\sigma i} d\varepsilon_{\sigma i} \times z_{\sigma i}}{\sum_{\sigma i} d\varepsilon_{\sigma i}}\right)_j \nonumber \\
&=& \left(\frac{\sum_{\sigma i} d\varepsilon_{\sigma i} \times (d\varepsilon_{\sigma i}/dm_j)}{\sum_{\sigma i} d\varepsilon_{\sigma i}}\right)_j.
\label{eq1_5}
\end{eqnarray}

The track averaged LET was calculated using the following formulation
\begin{eqnarray}
{\rm LET}_{t,j} &=& \frac{(\overline{dl \times y})_{j}}{(\overline{dl})_{j}} \nonumber \\
&=& \left(\frac{\sum_{\sigma i} d\varepsilon_{\sigma i}}{\sum_{\sigma i} dl_{\sigma i}}\right)_j.
\label{eq1_4}
\end{eqnarray}
Considering spectral density of the stepping-track length
\begin{eqnarray}
X_{\sigma i} = \frac{dl_{\sigma i}}{\sum_{\sigma i} dl_{\sigma i}},
\label{eq1_351}
\end{eqnarray}
we can easily show that ${\rm LET}_{t,j} = \left(\sum_{\sigma i} \frac{d\varepsilon_{\sigma i}}{dl_{\sigma i}} X_{\sigma i}\right)_j$.
The calculated dose, consistent with the distribution function used for track-averaging is simply sum over all energy transfers,
\begin{eqnarray}
{\rm D}_{t,j} &=& \sum_{\sigma i} z_{\sigma i} X_{\sigma i} \nonumber \\
&=& \left(\frac{\sum_{\sigma i} dl_{\sigma i} \times z_{\sigma i}}{\sum_{\sigma i} dl_{\sigma i}}\right)_j \nonumber \\
&=& \left(\frac{\sum_{\sigma i} dl_{\sigma i} \times (d\varepsilon_{\sigma i}/dm_j)}{\sum_{\sigma i} dl_{\sigma i}}\right)_j.
\label{eq1_5t}
\end{eqnarray}

Finally we consider a trivial distribution function represented by identity, $\mathbb{1}_{\sigma i}$.
In this representation,
${\rm D}_{j} = \sum_{\sigma i} z_{\sigma i} \mathbb{1}_{\sigma i} = \sum_{\sigma i} d\varepsilon_{\sigma i}/dm_j$ and  ${\rm LET}_{j}
= \left(\sum_{\sigma i} \frac{d\varepsilon_{\sigma i}}{dl_{\sigma i}} \mathbb{1}_{\sigma i}\right)_j = \sum_{\sigma i} d\varepsilon_{\sigma i}/dl_{\sigma i}$.

\section{Appendix C: Geant4 Monte Carlo simulations}
In Geant4 each pencil proton beams were simulated by irradiating a cylinder water phantom with 20 cm radius and 40 cm length. The mean deposited energies $\overline{d\varepsilon}$ and $\overline{(d\varepsilon)^2}$, the mean track length $\overline{dl}$ were scored within a linear array of voxels with 0.5 mm thickness.
Therefore $y_{1D}=\overline{(d\varepsilon)^2}/\overline{d\varepsilon}/\overline{dl}$ as well as other types of LET's were calculated.
The number of primary protons and the number of interactions per track were saved, in the same volume.  Then, the energy deposition, the track length, the number of primary proton and the number of interactions were accumulated, in each cell. All simulations used $10^6$ protons with a cut-off of 0.01 mm, where in Geant4, any particle with energy below the cut-off value assumed to not produce secondary particles anymore, and lose energy by the continuous slowing down approximation.
All simulation results presented used the QGSP-BIC-EMY physics list.
We used Gaussian proton energy spectrums with very small FWHM (0.18 MeV). Because of small divergence the simulated beam is mono-energetic.

{\bf Acknowledgement:}
The authors would like to acknowledge useful discussion and scientific exchanges with Drs., David Carlson, Fada Guan, Harald Paganetti and Robert Stewart.
The work at the University of Texas, MD Anderson Cancer Center was supported by the NIH / NCI under Grant No. U19 CA021239.




\noindent{\bf Authors contributions:}
RA: wrote the main manuscript, prepared figures, performed mathematical derivations and computational steps including Geant4 and Geant4-DNA Monte Carlo simulations and three dimensional surface fitting to the experimental data.
YH: contributed to Geant4 Monte Carlo simulations and writing the manuscript.
LB: provided experimental clonogenic cell survival data, aided in its interpretation and wrote the manuscript.
AC: wrote the main manuscript and contributed to computational analysis of the problem.
DG and RM: wrote the main manuscript and proposed scientific problem and co-supervised the project.

\noindent{\bf Competing financial interest:}
The authors declare no competing financial interests.

\noindent{\bf Corresponding Authors:}\\
$^\dagger$ ramin1.abolfath@gmail.com / ramin.abolfath@yale.edu \\
$^*$ rmohan@mdanderson.org


\end{document}